\documentclass{emulateapj}
\usepackage{times,color}

\usepackage{amsmath}

\shorttitle{FORMATION OF TURBULENT AND MAGNETIZED MOLECULAR CLOUDS}
\shortauthors{T. INOUE AND S. INUTSUKA}

\begin{document}

\title{
FORMATION OF TURBULENT AND MAGNETIZED MOLECULAR CLOUDS VIA ACCRETION FLOWS OF HI CLOUDS
}
\author{Tsuyoshi Inoue\altaffilmark{1} and Shu-ichiro Inutsuka\altaffilmark{2}}
\altaffiltext{1}{Department of Physics and Mathmatics, Aoyama Gakuin University, Fuchinobe, Chuou-ku, Sagamihara 252-5258, Japan; inouety@phys.aoyama.ac.jp}
\altaffiltext{2}{Department of Physics, Graduate School of Science, Nagoya University, Furo-cho, Chikusa-ku, Nagoya 464-8602, Japan}

\begin{abstract}
Using three-dimensional magnetohydrodynamic simulations including the effects of radiative cooling/heating, chemical reactions, and thermal conduction, we investigate the formation of molecular clouds in the multi-phase interstellar medium.
We consider the formation of molecular clouds due to accretion of H\textsc{i} clouds as suggested by recent observations.
Our simulations show that the initial H\textsc{i} medium is compressed and piled up behind the shock waves induced by accretion flows.
Since the initial medium is highly inhomogeneous as a consequence of the thermal instability, a newly-formed formed molecular cloud becomes very turbulent owing to the development of the Richtmyer-Meshkov instability.
The structure of the postshock region is composed of dense cold gas ($T<100$ K) and diffuse warm gas ($T>1,000$ K), which are spatially well mixed due to turbulence.
Because the energy source of the turbulence is the accretion flows, the turbulence becomes highly anisotropic biased towards the direction of the accretion flows.
The kinetic energy of the turbulence dominates the thermal, magnetic, and gravitational energies throughout the entire 10 Myr evolution.
However, the kinetic energy measured using CO-fraction-weighted densities is comparable to the other energies at $t\sim 5$ Myr, once the CO molecules are sufficiently formed owing to UV shielding.
This suggests that the true kinetic energy of turbulence in molecular clouds as a whole can be much larger than the kinetic energy of turbulence estimated using line-widths of molecular emission.
We find that clumps in a molecular cloud show a statistically homogeneous evolution as follows:
the typical plasma $\beta$ of the clumps is roughly constant $\langle \beta \rangle\simeq 0.4$;
the size-velocity dispersion relation is $\Delta v \simeq 1.5$ km s$^{-1}$ $(l/1\mbox{ pc})^{0.5}$, irrespective of the density;
the clumps evolve toward magnetically supercritical, gravitationally unstable cores;
the clumps seem to evolve into cores that satisfy the condition for fragmentation into binaries.
These statistical properties may represent the initial conditions of star formation.
\end{abstract}


\section{Introduction}
It is well known that molecular clouds are the sites of present-day star formation.
However, our understanding of the physical conditions of molecular clouds is very limited.
Investigations of the formation of molecular clouds should be one of the most promising way to reveal the physical conditions of molecular clouds.
Recent progress in numerical simulations of the interstellar medium has revealed many vital aspects of interstellar cloud formation.
Using one-dimensional hydrodynamics simulations, Hennebelle \& P\'erault (1999) and Koyama \& Inutsuka (2000) showed that shocked warm neutral medium (WNM or diffuse intercloud medium; $T\sim10^4$ K) evolves into cold neutral medium (CNM or interstellar cloud; $T\lesssim 10^2$ K) owing to radiative cooling.
In the follow-up two-dimensional simulation, Koyama \& Inutsuka (2002) found that the thermal instability (Field 1965, Field et al. 1969) that is caused by runaway cooling in shocked WNM generates highly clumpy small-scale clouds with velocity dispersion that are supersonic with respect to the CNM.

Since supersonic turbulence is a universal feature of interstellar clouds (Larson 1981, Heiles \& Troland 2005), the formation of turbulent clouds without ad-hoc external forcing is very advantageous to the study of the physical nature of clouds.
Hennebelle \& Audit (2007) and Hennebelle et al. (2007) showed that H\textsc{i} clouds formed by thermal instability can explain various observational characteristics.
In addition to the turbulence, magnetic fields are also an important ingredient of the interstellar clouds (Heiles \& Crutcher 2005).
Using magnetohydrodynamic (MHD) simulations, Inoue \& Inutsuka (2008, 2009) showed that molecular clouds can be formed behind a shock wave if the shock propagates almost parallel to the mean magnetic field, otherwise only H\textsc{i} clouds are formed since magnetic pressure prevents the accumulation of the gas (see also Hennebelle \& P\'erault 2000, Hartmann et al. 2001, Inoue et al. 2007).
Recently, employing three-dimensional MHD simulations, Hennebelle et al. (2008), Heitsch et al. (2009), Banerjee et al. (2009), and V\'azquez-Semadeni et al.(2011) studied the formation of molecular clouds due to converging flows of WNM along magnetic field lines.
Note that in the above-mentioned studies the gas is assumed to be atomic, and optically thin radiative cooling/heating are employed in the simulations.
Thus, the effects of chemical evolution including the effects of ultraviolet (UV) shielding and the resulting modifications to radiative cooling/heating mechanisms should be taken into account (e.g. Glover et al. 2007, 2010).

Recent observations of nearby galaxies, molecular clouds are suggested to be formed from H\textsc{i} clouds with $n\sim 10$ cm$^{-3}$ (Blitz et al. 2006, Fukui et al. 2009).
A detailed statistical study of molecular clouds in the Large Magellanic Cloud (LMC) showed that the timescale of molecular cloud evolution due to H\textsc{i} cloud accretion is approximately $\sim 10$ Myr (Fukui et al. 2008, Kawamura et al. 2009).
As discussed by Hartmann et al. (2001) and Inoue \& Inutsuka (2009), since the ISM can only be accumulated efficiently parallel to the mean magnetic field, forming a cloud with visual extinction $\langle A_{\rm v} \rangle>1$ ($N_{H}>2\times10^{21}$ cm$^{-3}$) by piling up the WNM ($n\sim 1$ cm$^{-3}$) takes $t_{\rm form} > 80\,\mbox{ Myr}$ $( A_{\rm v}/1 )$ $( v_{\rm flow}/8\,\mbox{km}\,\mbox{s}^{-1} )^{-1}$ $( n/1\,\mbox{cm}^{-3} )^{-1}$, even if the WNM is accumulated at the sound speed $v_{\rm flow}=c_{\rm s, wnm}\sim 8$ km s$^{-1}$.
This is an order of magnitude larger than the observational estimation of $\sim 10$ Myr.
Thus, in this paper, we examine the formation of molecular clouds by the accretion of H\textsc{i} clouds denser than the WNM,  as suggested by observations.

The plan of this paper is as follows:
in \S 2, we provide a possible scenario for molecular cloud formation and the numerical setup of our simulations;
the results of the simulations are shown in \S 3;
finally, in \S 4, we summarize our findings and discuss their implications.

\section{Formation Scenario and Numerical Setup}
\subsection{Scenario}
In Inoue \& Inutsuka (2008, 2009), we showed for the case in which the angle between the shock normal and the mean magnetic field is not parallel that the WNM compressed by an interstellar shock wave (such as a supernova blast wave, super-bubble, or galactic spiral shock) is cooled roughly isochorically and generates thermally unstable gas.
The resulting thermally unstable medium evolves into a complex of sheet-like H\textsc{i} clouds with $n\sim 30$ cm$^{-3}$ embedded in the diffuse WNM.
Since the total mass of H\textsc{i} clouds formed behind a super-bubble can be huge, they could be a mass reservoir of the building blocks of molecular clouds.
According to Inoue \& Inutsuka (2009), the average number density of the H\textsc{i} medium composed of H\textsc{i} clouds and WNM generated behind a shock is $\sim 10$ cm$^{-3}$, which agrees well with the observed H\textsc{i} mass reservoir of molecular clouds (Fukui et al. 2009, Dawson et al. 2011).

If the H\textsc{i} reservoir is swept up along the mean magnetic field by shock waves, e.g. due to a galactic spiral shock or super-bubble, molecular clouds can be formed.
In this case, because the mean density of the medium is an order of magnitude larger than the WNM, the timescale of mass accumulation to form a cloud of $\langle A_{\rm v} \rangle>1$ becomes $t_{\rm form} > 8\,\mbox{ Myr}$ $( A_{\rm v}/1 )$ $( v_{\rm flow}/8\,\mbox{km}\,\mbox{s}^{-1} )^{-1}$ $( n/10\,\mbox{cm}^{-3} )^{-1}$, consistent with the evolutionary timescale of observed molecular clouds (Kawamura et al. 2009).
In addition, if the H\textsc{i} reservoir is massive enough to be gravitationally unstable, the formation of molecular clouds due to gravitational contraction along mean magnetic field lines would also be possible.
Thus, in this paper, we examine the formation of molecular cloud due to the transonic accumulation of H\textsc{i} clouds along magnetic field lines.

\subsection{Formation of H\textsc{i} Reservoir}
In this section, we provide the computational setup for forming the H\textsc{i} reservoir for molecular cloud formation.
The detailed modeling of molecular cloud formation from this H\textsc{i} reservoir is presented in the next section.
The basic equations to be solved are the ideal MHD equations including the effects of chemical reactions, radiative cooling/heating, and thermal conduction.
$$\frac{\partial\,n_{\rm i}}{\partial t}+\vec{\nabla}\cdot(n_{\rm i}\,\vec{v})=f_{\rm i}(n_{\rm j},N_{\rm j},T,G_0)$$
$$\frac{\partial\,\rho\vec{v}}{\partial t}+\vec{\nabla}\cdot(p+\frac{B^2}{8\,\pi}+\rho\,\vec{v}\otimes\vec{v}-\frac{\vec{B}\otimes\vec{B}}{4\,\pi})=0,$$
$$\frac{\partial \,e}{\partial t}+\vec{\nabla}\cdot\{(e+p+\frac{B^2}{8\,\pi})\,\vec{v}-\frac{\vec{B}\cdot\vec{v}}{4\,\pi}\vec{B}\}\!=\!\vec{\nabla}\cdot\kappa\vec{\nabla} T-L(n_{\rm i},\,N_{\rm i},\,T,\,G_0),$$
$$\frac{\partial \,\vec{B}}{\partial t}=\vec{\nabla}\times(\vec{v}\times\vec{B}),$$\vspace{-0.25cm}
$$\rho=\sum_{\rm i} m_{\rm i}\,n_{\rm i},$$\vspace{-0.25cm}
$$e=\frac{p}{\gamma-1}+\frac{\rho\,v^2}{2}+\frac{B^2}{8\,\pi},$$
where the subscript $i$ denotes the chemical species p, H, H$_2$, He, He$^+$, C, C$^+$, and CO, $f_{\rm i}$ is the chemical reaction term for species $i$, and $L$ is the net cooling rate per unit volume.
The other symbols have their usual meaning.
The chemical reaction coefficients and the net cooling rate depend on temperature ($T$), the strength of the external radiation field ($G_0$), and the column densities of chemical species and dust ($N_{\rm i}$).
In this paper, we consider a requisite minimum set of chemical reactions and cooling/heating processes in order to follow the formation of an opaque molecular cloud from an optically thin H\textsc{i} medium.
We summarize the chemical reactions and cooling/heating processes in Tables 1 and 2, respectively.

The basic MHD equations are solved using a second-order Godunov-type finite volume scheme (van Leer 1979) that employs an approximate Riemann solver developed by Sano et al. (1999).
 The consistent method of characteristics with constrained transport is used for solving the induction equation (Clarke 1996, see also Evans \& Hawley 1988, Stone \& Norman 1992).
The integration of the ideal MHD part of the basic equations is done in a conservative fashion.
The cooling, heating, and thermal conduction terms are solved using the second-order explicit method.
We use the piecewise exact solution method developed by Inoue \& Inutsuka (2008), an unconditionally stable second-order method originally adopted for solving friction terms in two-fluid MHD equations, to calculate the chemical reaction terms.
We confirmed the accuracy of this method by solving the chemical reaction equations in one-zone test problems.
We find that the piecewise exact solution method is as accurate as the Crank-Nicolson scheme, even though it does not require iterative procedures or matrix inversion (Kamata 2012).
The timestep for the numerical integration is determined by the minimum timestep due to the CFL condition for the MHD fast mode, one-fifth of the local cooling time, and the stability condition for the parabolic conduction equation.

\begin{deluxetable*}{lll} \label{t1}
\tablewidth{0pt}
\tablecaption{Chemical Reactions}
\tablehead{Reaction &  Note  & Reference }
\startdata
Cosmic ray ionization & $\zeta_{\rm H}=3.0 \times10^{17}$ s$^{-1}$, $\zeta_{\rm He}=1.087\,\zeta_{\rm H}, \zeta_{\rm C}=3.846\,\zeta_{\rm H}$  &  1 \\
H$_{2}$ formation on grains & dust temperature $T_{\rm d}=10$ K is used & 2 \\
H$_{2}$ photo-dissociation & depends on UV strength $G_{0}$, column density $N_{\rm{H}_2}$, &  \\
& and visual extinction Av  & 3 \\
H$_{2}$ and CO dissociation & collisions with e, p, and H  & 4 \\
H$_{2}$ and CO dissociation & recombination of He$^+$  & 1 \\
H, He, and C ionization & collisions with e, p, H, and H$_{2}$ & 1,\,4\\
H$^{+}$, He$^{+}$, and C$^{+}$ recombination & --- & 4,\,5 \\
CO formation & depends on $G_0$ and Av & 6 \\
CO photo-dissociation  & depends on $G_0$, $N_{\rm{H}_2}$, $N_{\rm CO}$ and Av & 6,\,7 \\
C photo-ionization & depends on $G_0$, $N_{\rm C}$, $N_{\rm{H}_2}$ and Av & 2 \\
\enddata
\tablerefs{
(1) Millar et al. 1997;  (2) Tilelens \& Hollenbach 1985;  (3) Draine \& Bertoldi 1996; (4) Hollenbach \& McKee 1989; (5) Shapiro \& Kang 1987; (6) Nelson \& Langer 1997; (7) Lee et al. 1996
}
\end{deluxetable*}

\begin{deluxetable*}{lll} \label{t2}
\tablewidth{0pt}
\tablecaption{Cooling and Heating Processes}
\tablehead{Process &  Note  & Reference }
\startdata
Photoelectric heating by PAHs & depends on $G_0$ and Av &  1,\,2 \\
Cosmic ray heating & --- &  3 \\
H$_{2}$ photo-dissociative heating & depends on $G_0$, $N_{\rm{H}_2}$, and Av & 4 \\
Ly-$\alpha$ cooling & --- & 5 \\
C$^{+}$ cooling (158$\mu$m) & level population and escape probability &  \\
& as a function of $N_{\rm{C}^+}$ are taken into account & 6 \\
O cooling (63$\mu$m) & escape probability as a function of $N_{\rm{O}}=x_{\rm O}\,N_{\rm tot}$ &  \\
&  is taken into account ($x_{\rm O}=3.2\times10^{-4}$). & 2,\,6 \\
CO ro-vibrational cooling & escape probability as a function of $N_{\rm{CO}}$ & \\
& is taken into account & 7,\,8,\,9\\
Cooling due to rec. of  \\
electrons with grains \& PAHs & --- & 1 \\
Thermal conduction & collisions of H and H$_2$ & 10 \\
\enddata
\tablerefs{
(1) Bakes \& Tielens 1994; (2) Wolfire et al. 2003; (3) Goldsmith \& Langer 1978; (4) Black \& Dalgano 1977; (5) Spitzer 1978; (6) de Jong et al. 1980; (7) Hollenbach \& MacKee 1979; (8) Hosokawa \& Inutsuka 2006; (9) Hollenbach \& McKee 1989; (10) Parker 1969
}
\end{deluxetable*}

\begin{figure}[t]
\epsscale{1.}
\plotone{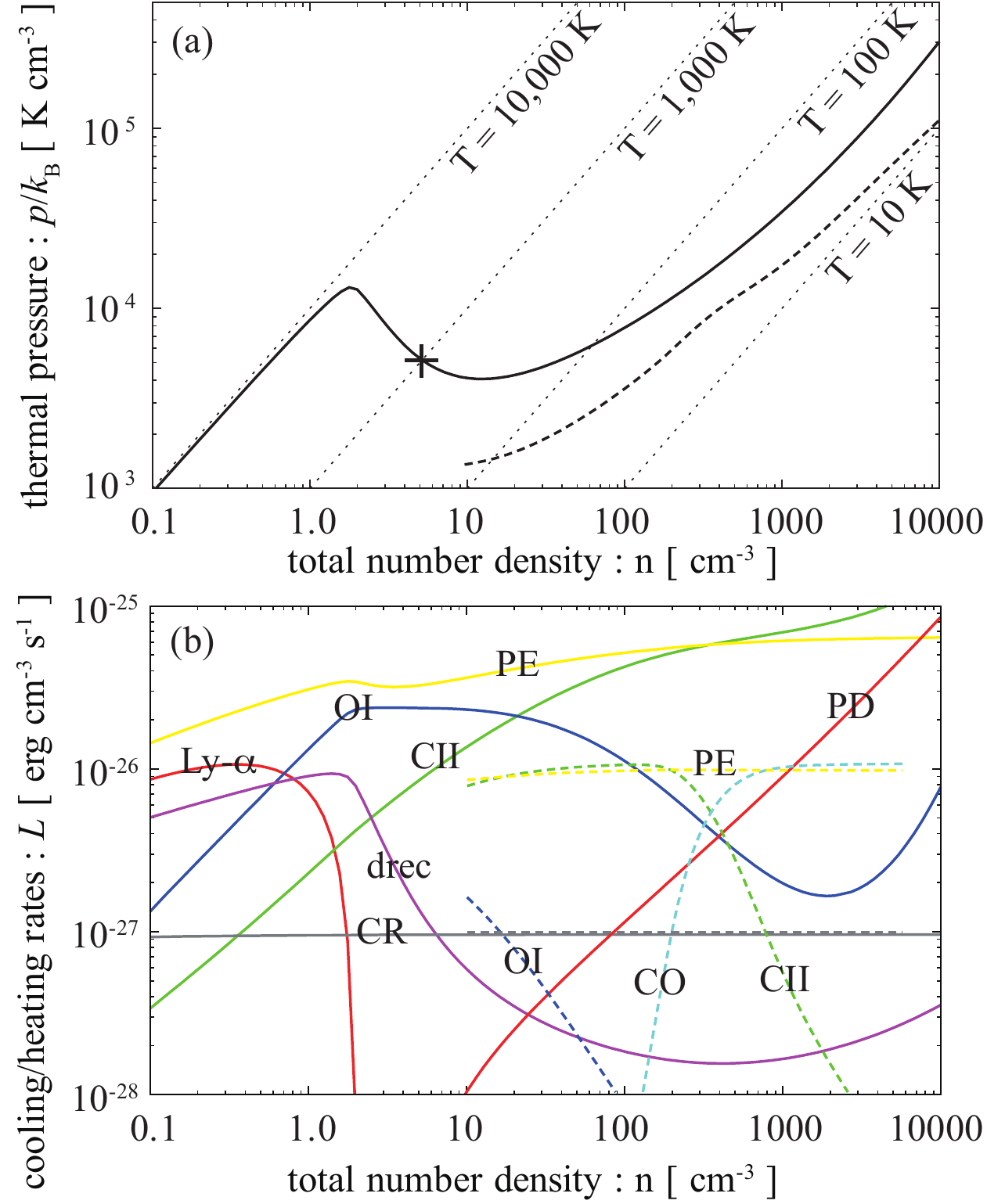}
\caption{
Panel (a): Thermal and chemical equilibrium curve for media of Av=0 (optically thin, {\it solid}) and Av=1 ({\it dashed}).
Dotted lines are isotherms of $T=10^1$, $10^2$, $10^3$, and $10^4$ K.
Panel (b): Cooling and heating rates of the thermal and chemical equilibrium states for Av=0 (optically thin, {\it solid}) and Av=1 media ({\it dashed}).
The symbol Ly-$\alpha$ denotes Ly-$\alpha$ cooling, CII denotes C$_{\rm II}$ fine structure line (158$\mu$m) cooling, OI denotes O$_{\rm I}$ fine structure line (63$\mu$m) cooling, DREC denotes the effect of cooling due to the recombination of electrons with grains and PAHs, CO denotes ro-vibrational cooling by CO molecules, PE denotes photo-electric heating by PAHs, CR denotes cosmic-ray heating, and PD denotes photo-dissociative heating of H$_2$ molecules.
}
\label{f1}
\end{figure}

We prepare an H\textsc{i} medium composed of H\textsc{i} clouds and WNM as building blocks of a molecular cloud.
In this stage, we assume optically thin cooling, heating and chemical reactions, i.e., the effects of UV shielding in the chemical reaction and heating/cooling terms are switched off.
The background UV field strength is set to the Habing flux $G_0=1$ ($=1.6\times10^{-3}$ erg cm$^{-2}$ s$^{-1}$).
The thermal and chemical equilibrium state of this optically thin medium is shown as a solid line in the number density-pressure plane in panel (a) of Figure~\ref{f1}.
We setup an initial state of thermally unstable gas with density fluctuations, the spectrum of which is a power law with the Kolmogorov spectral index ($P_{\rm 3D}\propto k^{-11/3}$).
Thermal pressure and average number density are set to $p/k_{\rm B}=5.2\times10^3$ K cm$^{-3}$ and $\langle n \rangle=\sum\langle n_{\rm i}\rangle=5.2$ cm$^{-3}$, where the initial abundances are $x_{\rm H}\equiv n_{\rm H}/\sum n_{\rm i}=0.91,\,x_{\rm p}=1.7\times10^{-3},\,x_{\rm H_{2}}=1.3\times10^{-6},\,x_{\rm He}=0.090,\,x_{\rm He^+}=1.3\times10^{-4},\,x_{\rm C^+}=1.4 \times10^{-4},\,x_{\rm C}=6.0 \times10^{-9}, $ and $\,x_{\rm CO}=5.7 \times10^{-18}$.
The initial mean molar weight and the total mass in the numerical domain are $m_{\rm mean}=1.27\,m_{\rm p}$ and $M_{\rm tot}=1305\,M_{\rm sol}$, respectively.
The mean density and pressure of this initial state is shown as a cross in Figure~\ref{f1}(a).
We use a cubic numerical domain of (20 pc)$^3$ in volume which we divide into $1024^3$ uniform cells, such that the numerical resolution is $\sim 0.02$ pc.
The initial medium is uniformly magnetized by a field of strength 5.0 $\mu$G oriented in the $+x$-direction.
As shown by Inoue \& Inutsuka (2009), such a thermally unstable medium can be formed behind a shock wave that propagates in turbulent WNM. 
Periodic boundary conditions are used for this stage of H\textsc{i} cloud formation.

\begin{figure}[t]
\epsscale{1.}
\plotone{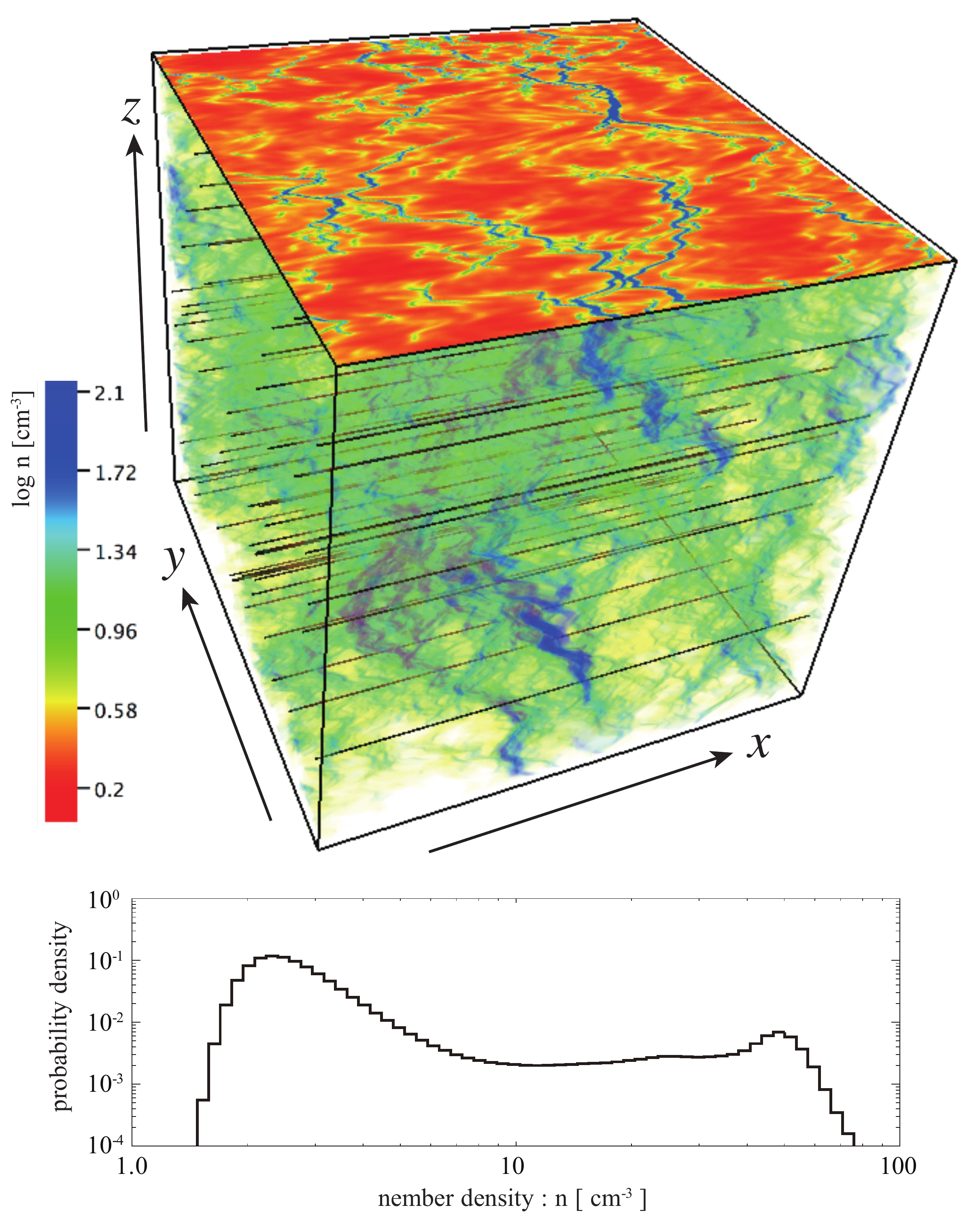}
\caption{
Volume rendering map of the density resulting from 8 Myr integration (several cooling times for the initial thermally unstable gas).
Regions in blue are H\textsc{i} clouds with $n\ge 10$ cm$^{-3}$ and $T\sim 100$ K that are formed as a consequence of the thermal instability.
Magnetic field lines are shown as black lines.
A two-dimensional slice is shown as the top surface in which red represents $n<3$ cm$^{-3}$, yellow and green represent $3$ cm$^{-3}<n<10$ cm$-3$, and blue represents $n>10$ cm$^{-3}$.
The PDF of the number density is plotted in the bottom panel.
}
\label{f2}
\end{figure}

The three-dimensional structure of the density resulting from 8 Myr of integration (several cooling times for the initial thermally unstable gas) is shown in the top panel of Figure~\ref{f2}.
Regions in blue are H\textsc{i} clouds with $n\ge 10$ cm$^{-3}$ and $T\sim 100$ K that are formed as a consequence of the thermal instability.
Magnetic field lines are shown as black lines.
A two-dimensional slice is also shown, in which regions in red represent $n<3$ cm$^{-3}$, regions in yellow and green represent $3$ cm$^{-3}<n<10$ cm$-3$, and regions in blue represent $n>10$ cm$^{-3}$.
The probability distribution function (PDF) of the number density is plotted in the bottom panel of Figure~\ref{f2}.
Condensation driven by the thermal instability arises along the magnetic field lines, since condensation perpendicular to the field is suppressed due to the enhancement of magnetic pressure (Hennebelle \& P\'erault 2000, Inoue et al. 2007, Inoue \& Inutsuka 2008, 2009).
This results in the formation of sheet-like clouds.
Note that the gas in the intercloud medium is also thermally unstable due to runaway heating that causes it to evolve towards the WNM phase.
Linear stability analyses of the thermal transition layer between an H\textsc{i} cloud and WNM have shown that the transition layer is always unstable with respect to corrugational perturbations (Inoue et al. 2006, Stone \& Zweibel 2009).
Thus, the resultant H\textsc{i} clouds are highly corrugated.
The sheet-like morphology of the H\textsc{i} clouds, their magnetization ($\sim 5\,\mu$G), typical density ($n\sim 50$ cm$^{-3}$), and temperature ($T\sim 100$ K) agree well with observed H\textsc{i} clouds in the ISM (Heiles \& Troland 2003).
In addition, the mean density of the H\textsc{i} reservoir composed of H\textsc{i} clouds and the WNM is also consistent with observed H\textsc{i} reservoirs (Fukui et al. 2009).
As we will show in \S 3, the mass spectrum of the H\textsc{i} clouds is fitted by a power-law of $dN(M)/dM\propto M^{-1.8}$ that shows good agreement with the theoretical mass spectrum of H\textsc{i} clouds formed as a result of the thermal instability (Hennebelle \& Audit 2007, Hennebelle \& Chabrier 2008).

\subsection{Molecular Cloud Formation}
In this section, we provide our numerical setup for molecular cloud formation.
We use the H\textsc{i} medium obtained in the previous section as an initial condition, and consider the converging flows of the H\textsc{i} reservoir along the mean magnetic field ($\pm x$-direction).
For this purpose, we set the initial velocity to $v_x=v_{\rm cnv}\,\tanh (-x/ 1\,{\rm pc})$, where the center of the numerical domain is defined as $x=y=z=0$.
In this paper, we examine the case of a converging flow velocity of $v_{\rm cnv}=20$ km s$^{-1}$, comparable to the case where H\textsc{i} medium is swept by a galactic spiral shock or by collisions of H\textsc{i} shells in super-bubbles.
We use periodic boundary condition at the $y$- and $z$-boundary planes.
The initial setting is also periodic in the $x$-direction except for $v_x$ given above.
In effect, we impose continuous flows of the initial two-phase medium at the $x$-boundary planes at $x=-10$ pc and $10$ pc. 
Thus, there is a continuous mass input into the numerical domain from the $x$-boundaries.
In this paper, we omit the dynamical effects of self-gravity, but we estimate it using a posteriori calculations in \S 3.2 and \S3.5. 
The neglect of self-gravity for the duration of this simulation ($t=10.0$ Myr) is justified, because the gravitational energy does not become the dominant component of the energy budget (see \S 3.2) and most of the dense clumps in the molecular cloud stay in a gravitationally unbound state (see \S 3.5).

Since the converging flow velocity is a few times larger than the sound speed in the WNM ($c_{\rm s, w}\sim 10$ km s$^{-1}$) and much larger than that in the CNM ($\sim 1$ km s$^{-1}$), two shock waves that bound the shocked slab are formed.
In order to consider the effects of ultra-violet (UV) photon shielding and the resulting formation of molecules and the transition of the cooling/heating processes, we take into account the dependence of the local column densities ($N_{\rm i}$) on the chemical reaction coefficients and cooling/heating rates (see Tables 1 and 2).
In this paper, we assume that the sources of the UV field, which dissociates the molecules and causes photoelectric heating by PAHs, are located far from the region of molecular cloud formation.
Because of the periodic conditions for $y$/$z$ boundaries, the global morphology of the generating molecular cloud should be regarded as sheet-like structure that lies approximately in the $y$-$z$ plane.
Thus, the UV photons propagating along the $y$-$z$ plane are well shielded in contrast to those propagating along the $x$ direction.
Hence, we use a two ray approximation for UV field that irradiates the numerical domain from the $x=\pm10$ pc planes toward the $\mp x$ directions with strength $G_0=0.5$ in both rays.
Note that the shielding factor of the UV flux is larger for an isotropic background UV radiation field, since the shielding increases with increased column density in the pathway of UV photons if the angle between a UV ray and the x-axis becomes larger.
This means that the formation of molecules is easier for an isotropic background UV radiation field with the same mean intensity.
Thus, the two-ray approximation considered in this analysis gives the upper bound for the formation timescale of molecules.
We also calculate the approximate escape probabilities for the emission of C$_{\rm II}$/O$_{\rm I}$ fine structure lines and CO ro-vibrational lines using column densities along $\pm x$ directions, because the medium is opaque in the $y$- and $z$-directions.
The local chemical reaction coefficients (cooling/heating rates) are modified from the optically thin case by multiplying them by the shielding factors (escape probabilities) that are calculated by averaging the two shielding factors for the right and left rays (two escape probabilities for the left and right directions).
Their specific forms, which were originally formulated in a slab geometry, are found in the references in Tables 1 and 2.

In Figure~\ref{f1}(a), we plot the thermal and chemical equilibrium state under the assumed visual extinction of Av=1 $(N_{\rm tot}\equiv N_{\rm H}+2\,N_{\rm H_{2}}+N_{\rm p}+N_{\rm He}+N_{\rm He^+}=1.9\times10^{21}\,\rm{cm}^{-2})$ as a dashed line.
In panel (b) of Figure~\ref{f1}, we also plot the cooling and heating rates of the thermal and chemical equilibrium states for the Av=0 (optically thin) and Av=1 cases as solid and dashed lines, respectively.
These panels shows that our numerical code can adequately treat media of warm atomic gas of $T\sim10^4$ K to cold molecular gas of $T\sim 10$ K.

\section{Results}
\subsection{Formation of a Molecular Cloud}
\begin{figure}[t]
\epsscale{1.}
\plotone{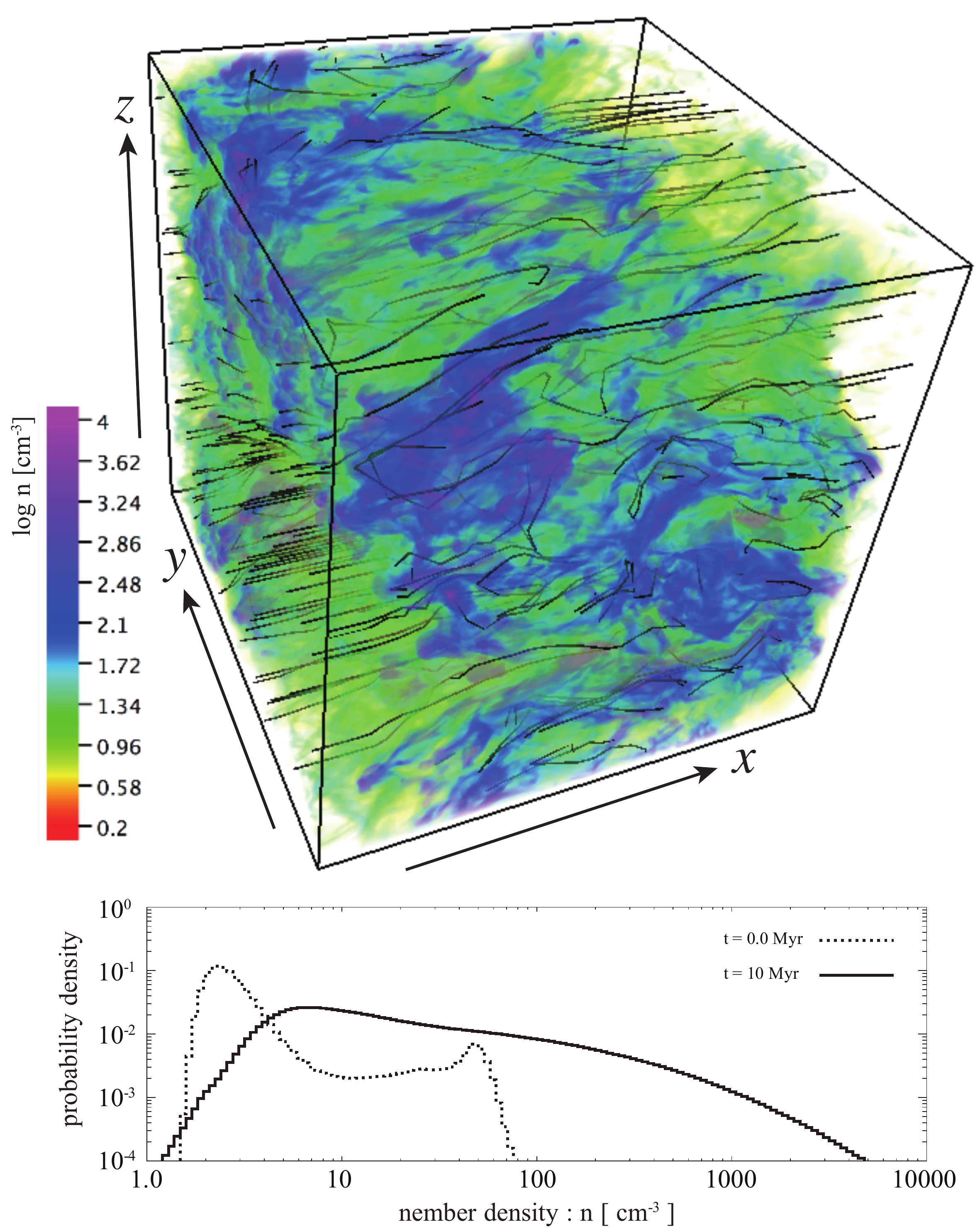}
\caption{
Volume rendering map of the density at $t=10.0$ Myr after the onset of converging flows ({\it top}).
Regions in yellow and green represent 3 cm$^{-3}$ $<n<30$ cm$^{-3}$ that correspond to preshock H\textsc{i} clouds and shocked WNM, regions in blue represent $10^2$ cm$^{-3}$ $<n<10^3$ cm$^{-3}$ that correspond to shocked H\textsc{i} clouds and molecular clumps, and regions in magenta represent dense clumps of $n>10^3$ cm$^{-3}$ that are formed, e.g., by collisions of clumps.
Magnetic field lines are shown as black lines.
The PDF of the number density ({\it solid}) and that of the initial H\textsc{i} medium ({\it dotted}) are plotted in the bottom panel.
The PDF at $t=10$ Myr is calculated only for gas in the region bounded by two accretion shocks.
}
\label{f3}
\end{figure}

\begin{figure}[t]
\epsscale{1.}
\plotone{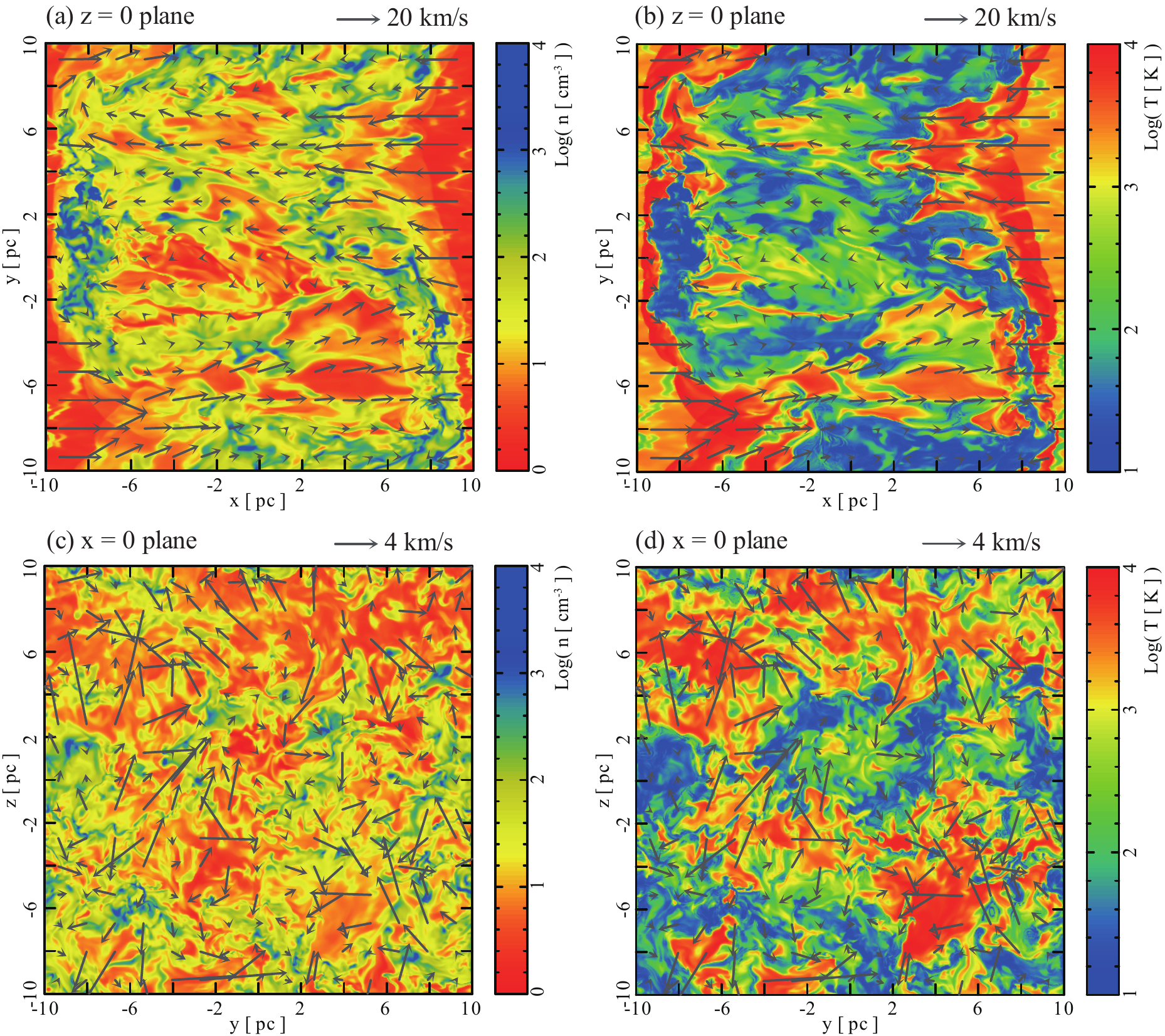}
\caption{
Panel (a): cross section map of number density in the $z=0$ plane.
Panel (b): temperature map in the $z=0$ plane.
Panel (c): number density map in the $x=0$ plane.
Panel (d): temperature map in the $x=0$ plane.
All the maps are snapshots at $t=10.0$ Myr.
Arrows represent the projected velocity on each plane.
}
\label{f4}
\end{figure}

The converging H\textsc{i} flows induce two accretion shock waves that bound a highly inhomogeneous cloud.
Owing to continuous mass accretion, the mean column density of the cloud along the $x$-axis increases linearly with time and a molecular cloud is formed.
In the top panel of Figure~\ref{f3}, we show the three-dimensional density map of the numerical domain at $t=10.0$ Myr after the onset of the converging flows.
Regions in yellow and green represent 3 cm$^{-3}$ $<n<30$ cm$^{-3}$ that correspond to the preshock H\textsc{i} clouds and shocked WNM, regions in blue represent $10^2$ cm$^{-3}$ $<n<10^3$ cm$^{-3}$ that correspond to the shocked H\textsc{i} clouds and molecular clumps, and regions in magenta represent dense clumps of $n>10^3$ cm$^{-3}$ that are formed, e.g., by collisions of clumps.
The PDFs of the number density ({\it solid}) and the initial HI medium ({\it dotted}) are plotted in the bottom panel.
The PDF at $t=10$ Myr is calculated only for gas in the region bound by the two accretion shocks.
We show cross-sectional maps of the number density ($n\equiv \sum n_{\rm i}$) and temperature in the $z=0$ and $x=0$ planes in Figure~\ref{f4}, where arrows represent the projected velocity on each plane.

As shown by Koyama \& Inutsuka (2002), even if the preshock medium is composed of WNM alone, nonlinear evolution of the thermal instability leads to the formation of turbulent cloud composed of small-scale cloudlets.
In the present case, because the accreting two-phase medium is much more inhomogeneous than the single-phase WNM, the generated structure of the cloud is even more turbulent and inhomogeneous.
This is because the interaction of the accretion shock waves with upstream large density inhomogeneities triggers the Richtmyer-Meshkov instability (see, e.g., Nishihara et al. 2010 for review) that induce rippling in the shock front.
As a result, turbulent vorticity is created behind the curved shock as a consequence of Crocco's theorem (Kida \& Orszag 1990; Kevlahan \& Pudritz 2009; see also Inoue et al. 2009, 2010, 2012 for recent simulations of the shock-cloud interaction).
In the panel (b) of Figure~\ref{f4}, it is evident that the two accretion shock fronts have became curved due to the Richtmyer-Meshkov instability.
Note that the growth timescale of the Richtmyer-Meshkov instability is given by the shock crossing time of the accreting CNM $\sim 1$ pc$/20$ km s$^{-1}$ $\sim 0.05$ Myr, so it is influential from the very early stage of cloud formation.

\begin{figure}[t]
\epsscale{0.8}
\plotone{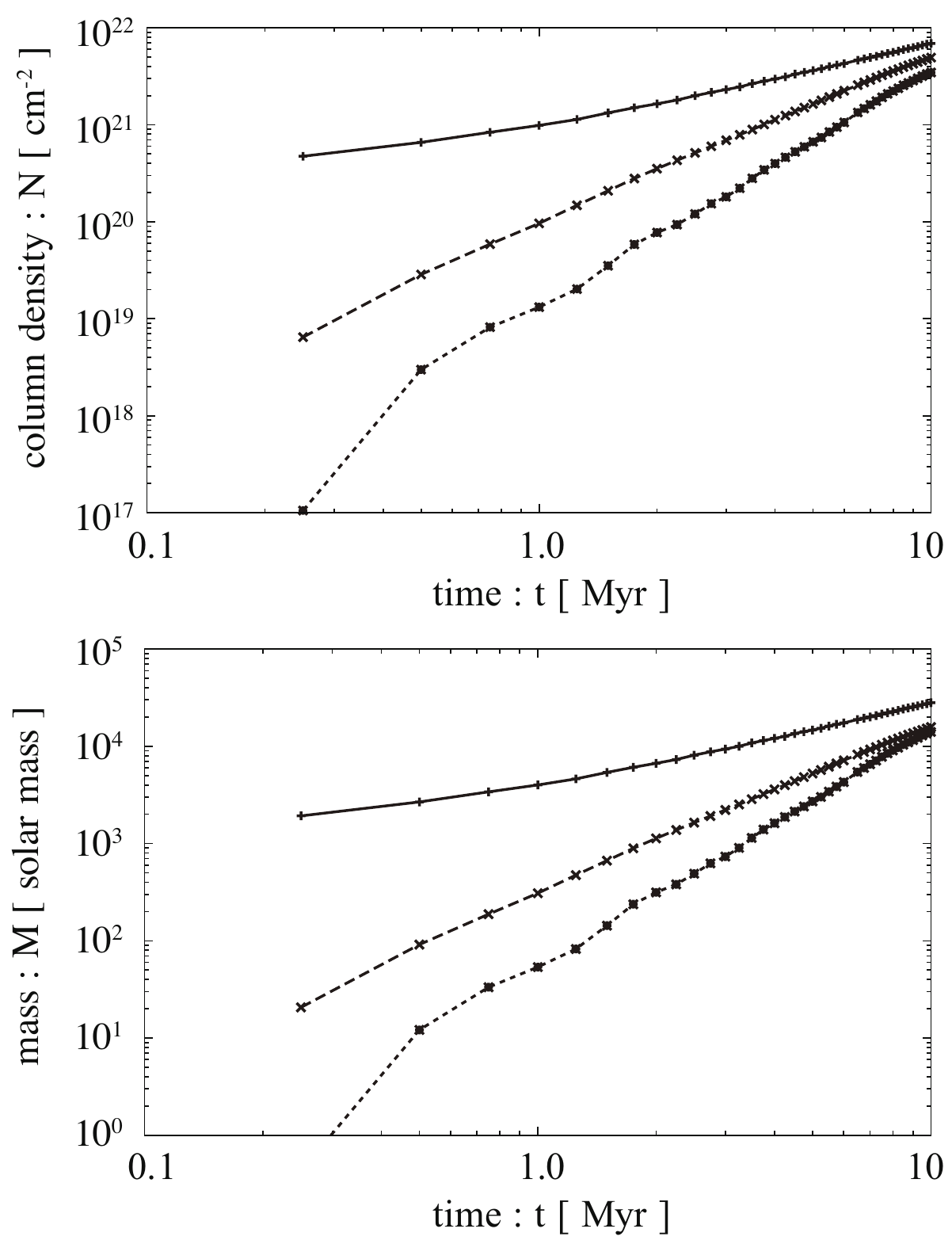}
\caption{
Top: Average column densities along $x$-axis for H and He atoms ($N_{\rm H}=\langle \int [n_{\rm p}+n_{\rm H}+2\,n_{{\rm H}_2}+n_{\rm He}+n_{{\rm He}^+}] dx\rangle_{y,z}$; {\it solid}), for H$_{2}$ molecules ($N_{{\rm H}_2}=\langle \int n_{{\rm H}_2} dx\rangle_{y,z}$; {\it dashed}), and for CO-fraction-weighted density ($N_{\rm CO}=\langle \int f_{\rm CO}\,[n_{\rm p}+n_{\rm H}+2\,n_{{\rm H}_2}+n_{\rm He}+n_{{\rm He}^+}] dx\rangle_{y,z}$; {\it dotted}), where $f_{\rm CO}\equiv n_{\rm CO}/(n_{{\rm C}^+}+n_{\rm C}+n_{\rm CO})$.
Bottom: Evolution of total mass ($M_{\rm tot}=\int \rho\,dV$; {\it solid}), H$_2$ mass ($M_{{\rm H}_2}=\int 2\,m_{\rm p}\,n_{{\rm H}_2} dV$; {\it dashed}), and CO-fraction-weighted total mass ($M_{\rm CO}=\int \,f_{\rm CO}\,\rho\,dV$; {\it dotted}).
}
\label{f5}
\end{figure}

In the top panel of Figure~\ref{f5}, we plot the evolution of the average column densities along the $x$-axis for H and He atoms ($N_{\rm H}=\langle \int [n_{\rm p}+n_{\rm H}+2\,n_{{\rm H}_2}+n_{\rm He}+n_{{\rm He}^+}] dx\rangle_{y,z}$; {\it solid}), for H$_{2}$ molecules ($N_{{\rm H}_2}=\langle \int n_{{\rm H}_2} dx\rangle_{y,z}$; {\it dashed}), and for the CO-fraction-weighted density ($N_{\rm CO}=\langle \int f_{\rm CO}\,[n_{\rm p}+n_{\rm H}+2\,n_{{\rm H}_2}+n_{\rm He}+n_{{\rm He}^+}] dx\rangle_{y,z}$; {\it dotted}), where $f_{\rm CO}\equiv n_{\rm CO}/(n_{{\rm C}^+}+n_{\rm C}+n_{\rm CO})$.
In the bottom panel, we plot the evolution of total mass ($M_{\rm tot}=\int \rho\,dV$; {\it solid}), H$_2$ mass ($M_{{\rm H}_2}=\int 2\,m_{\rm p}\,n_{{\rm H}_2} dV$; {\it dashed}), and CO-fraction-weighted total mass ($M_{\rm CO}=\int \,f_{\rm CO}\,\rho\,dV$; {\it dotted}) in the numerical domain.
Because self-shielding of UV is effective for H$_2$, whereas UV shielding by dust is necessary for CO formation, the H$_2$ mass and column density is always larger than the CO-fraction-weighted mass and column density.
At $t\gtrsim 5$ Myr, the total mass and the molecular mass measured by CO becomes of the same order of magnitude.
Thus, in the present case, the timescale at which most of C nuclei in the cloud become molecule is $5$-$10$ Myr, which is consistent with the estimated timescale of molecular cloud formation associated with super-shells (Dawson et al. 2011).

It has been pointed out that, in a cloud formed under the influence of the thermal instability, the cloud is not composed only of cold gas, but also the fragmented clumps of clouds embedded in warm diffuse gas ($T\gtrsim1000$ K and $n\lesssim10$ cm$^{-3}$; e.g., Koyama \& Inutsuka 2002, Hennebelle et al. 2008).
Note that the diffuse warm gas is the thermally unstable shocked WNM that eventually evolve into cold clumps due to the cooling (Banerjee et al. 2009).
In previous studies of molecular cloud formation by converging WNM flows, the simulations are performed assuming optically thin radiative cooling and heating.
Thus, the warm gas was irradiated by an overestimated UV field that led to excessive photoelectric heating and thus lengthened the cooling time and the lifetime of shocked warm gas.
In the present case, however, even after the formation of molecules, i.e., even in the medium where the UV field is shielded, the shocked warm gas between the cold clumps survives as seen in Figure~\ref{f4}.
The volume fraction of diffuse gas with $T>10^3$ K ($\langle n\rangle_{T>10^3\,\rm{K}}=8.44$ cm$^{-3}$) in the region bounded by the two accretion shock waves at $t=10.0$ Myr is 51.5\%!
The reason for the survival of this diffuse gas is explained as follows:
The diffuse gas (more precisely the shocked WNM) is continuously supplied by the two accretion shocks.
Since the typical cooling time for the shocked diffuse gas is only a few Myr, it seems very hard for the diffuse gas to exist deep inside the molecular cloud.
However, as we will show in more depth in \S3.3, the shock bounded region is highly turbulent and the velocity dispersion for the diffuse component of $T>10^3$ K is $\Delta v\gtrsim$ 10 km s$^{-1}$.
Thus, the turbulent mixing length is evaluated as $l_{\rm mix}\sim \Delta v\,t_{\rm cool}\gtrsim 10$ pc, which makes the supply of diffuse gas to deep inside the molecular cloud possible before the gas is cools down.
Note that, as discussed in Hennebelle \& Inutsuka (2007), heating due to ambipolar diffusion makes the net cooling time longer than in the present case and would further increase the lifetime of diffuse gas in realistic molecular clouds (see also, Inoue et al. 2007 and Vasquez-Semadeni et al. 2011 for the effects of ambipolar diffusion on cloud formation).

\begin{figure}[t]
\epsscale{1.}
\plotone{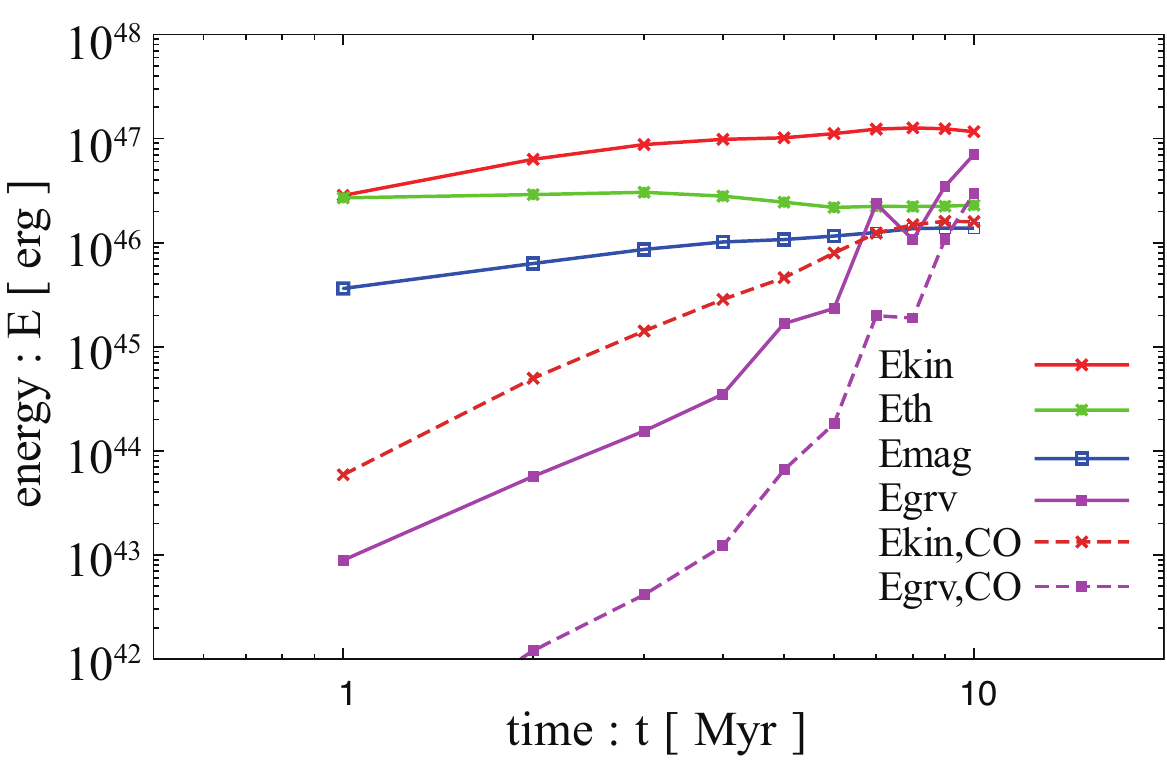}
\caption{
Evolution of the kinetic ({\it red}), magnetic ({\it blue}), thermal ({\it green}), and gravitational ({\it magenta}) energies.
Kinetic and gravitational energies calculated using the CO-fraction-weighted density field are also plotted as dashed red and dashed magenta lines, respectively.
The energies are calculated only for the gas in the region bounded by the two accretion shock waves, i.e., energies in the unshocked converging flows are excluded.
}
\label{f6}
\end{figure}

\subsection{Evolution of Global Quantities}
In Figure~\ref{f6}, we plot the evolution of the kinetic, thermal, magnetic, and gravitational energies as solid lines in red, green, blue, and magenta, respectively.
We also plot the kinetic and gravitational energies using the CO-fraction-weighted density field as dashed red and dashed magenta lines, respectively.
These energies are calculated only for gas in the region bounded by the two accretion shock waves, i.e., the energies in the unshocked converging flows are excluded.
The positions of the two accretion shocks ($x_{\rm sh}[y,z]$) are identified by searching for the maximum and minimum $x$-coordinate values where the thermal pressure is larger than $p_{\rm th}=10^4\,k_{\rm B}$ erg cm$^{-3}$, since the converging H\textsc{i} medium is almost isobaric with $\langle p\rangle=5\times 10^3\,k_{\rm B}$ erg cm$^{-3}$.
The gravitational energy ($\int \rho\,\{ \vec{r}-\vec{r}_{\rm c} \}\cdot \vec{\nabla}\psi\,dV$) is calculated using the numerical solution of the Poisson equation ($\Delta \psi =4\pi\,G\,\rho$, and $4\pi\,G\,\rho\,f_{\rm CO}$ for CO-fraction-weighted one) by assuming vacuum state outside the boundaries in $x$ direction and periodic boundary conditions for the $y$- and $z$-directions.
The gravitational energies calculated for the raw density field and for the CO-fraction-weighted field become comparable (same order) at $t\sim7$ Myr, which is consistent with the result shown in Figure~\ref{f5} that the total mass and the CO-fraction-weighted mass become the same order at $t \sim 5$-$10$ Myr.
The epochs at which the gravitational energy and turbulent energy measured by the CO-fraction-weighted data become comparable are approximately the same ($t\sim 7$ Myr).
This indicates that the scenario of molecular cloud formation examined in this paper is consistent with the fact that most observed molecular clouds are in an approximately virial equilibrium state.

As discussed in the previous section, the resultant molecular cloud is composed of dense cold clumps embedded in diffuse warm gas.
Figure~\ref{f6} shows that the total kinetic energy in the shock bounded region ({\it solid red}) is an order of magnitude larger than that contained in the molecular clumps ({\it dotted red}).
This is because the velocity dispersion of the diffuse gas component is an order of magnitude larger than that of the molecular gas, while the density of the molecular gas is only an order of magnitude larger than the atomic gas and the volume filling fractions of the atomic and molecular components are comparable.
For instance, in the shock bounded region at $t=10.0$ Myr, the average density and velocity dispersion in gas with CO fraction $f_{\rm CO}$ is larger than 0.1 are $\langle n \rangle_{f_{\rm CO}>0.1}=383$ cm$^{-3}$ and $\Delta v_{f_{\rm CO}>0.1}=2.86$ km s$^{-1}$, and in atomic gas with $f_{\rm CO}<0.1$ they are $\langle n \rangle_{f_{\rm CO}<0.1}=35.0$ cm$^{-3}$ and $\Delta v_{f_{\rm CO}>0.1}=12.2$ km s$^{-1}$.
This suggests that the true turbulent kinetic energy in the cloud as a whole can be much larger than the kinetic energy of turbulence estimated using line-widths of molecular emissions.

\begin{figure}[t]
\epsscale{1.}
\plotone{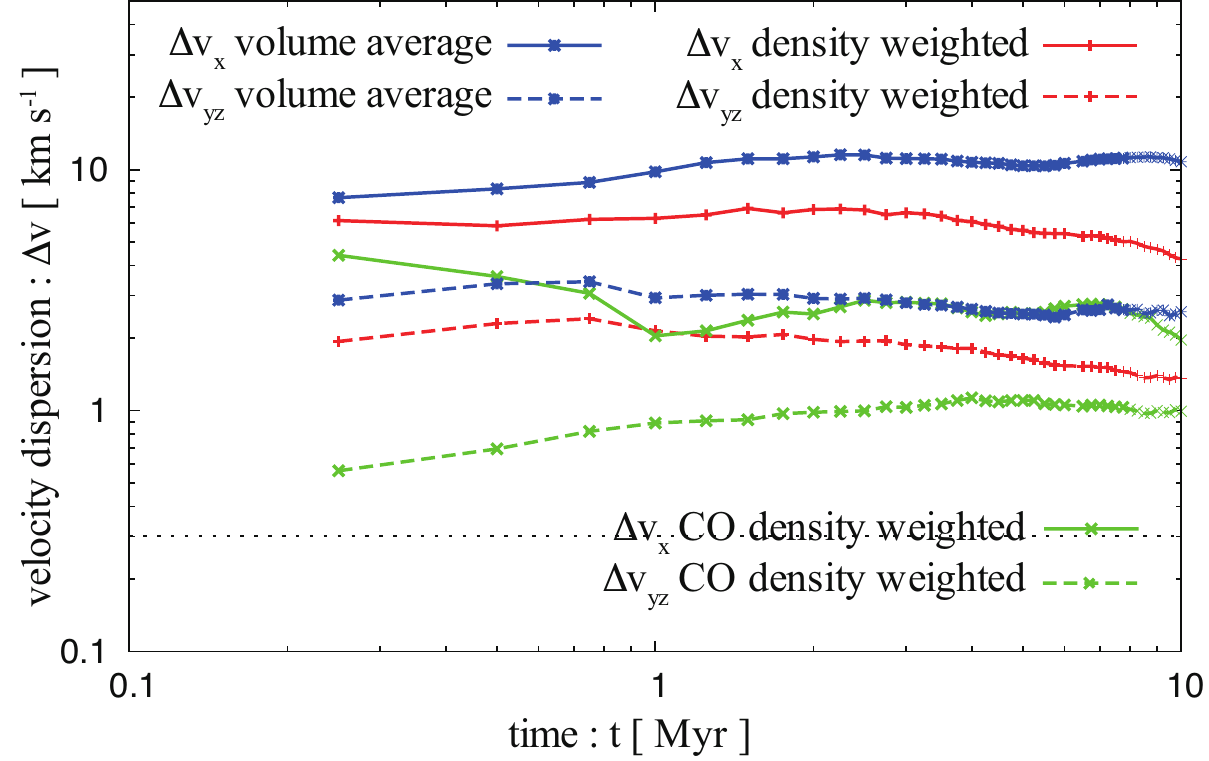}
\caption{
Evolution of the density-weighted ({\it red}), CO-density-weighted ({\it green}), and volume-averaged ({\it blue}) velocity dispersions in the shock bounded region.
Solid lines represent the $x$ component velocity dispersion ($\Delta v_x=\{\langle v_x- \langle v_x \rangle\}^2 \rangle^{1/2}$), dashed lines represent that of the $y$ and $z$ component average ($\{\Delta v_y^2+\Delta v_z^2\}^{1/2}/2$).
A typical sound speed in molecular gas, $c_{\rm s}\simeq 0.3\,(T/20\,\mbox{K})^{1/2}$ km s$^{-1}$, is also plotted as a dashed line, where the temperature $T=20$ K is the average CO-density-weighted temperature.
}
\label{f7}
\end{figure}

\subsection{Turbulence}
\begin{figure}[t]
\epsscale{0.9}
\plotone{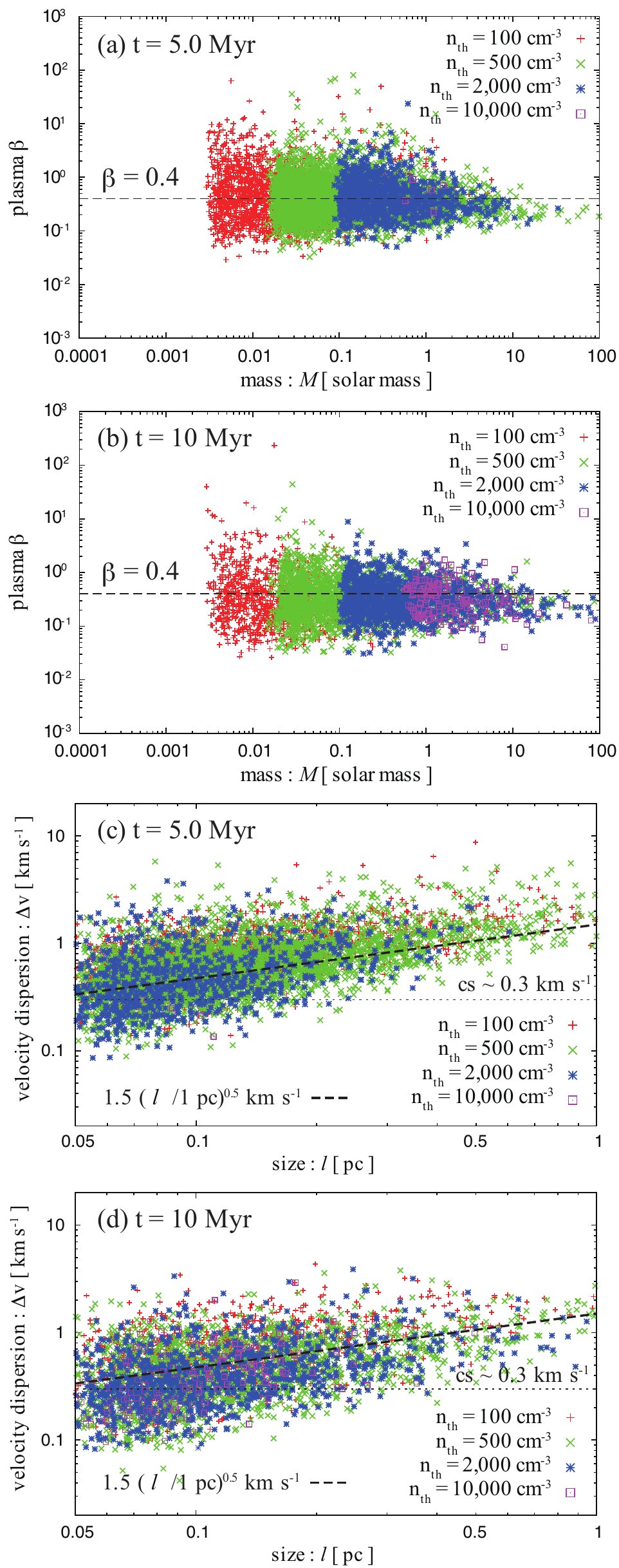}
\caption{
Scatter plots of the plasma $\beta$ ($\langle 8\,\pi\,p/B^2 \rangle$) as a function of clump mass and the velocity dispersion as a function of clump scale with threshold densities of $n_{\rm th}=100$ cm$^{-3}$, 500 cm$^{-3}$, 2,000 cm$^{-3}$, and 10,000 cm$^{-3}$.
Panels (a) and (b) show the plasma $\beta$ at $t=5.0$ Myr and $t=10$ Myr, respectively, and panels (c) and (d) show the size-velocity dispersion relation at $t=5.0$ Myr and $t=10$ Myr, respectively.
The size of the clumps is evaluated as $l=\sqrt{I_{1}/M}$, where $I_{1}$ is the largest eigenvalue of the inertial matrix and $M$ is the mass of the clump.
}
\label{f9}
\end{figure}

In Figure~\ref{f7} we plot the evolution of the density-weighted, CO-density-weighted, and volume averaged velocity dispersions in the shock bounded region as red, blue, and green lines, respectively.
Solid lines represent the $x$-component velocity dispersion ($\Delta v_x=\{\langle v_x- \langle v_x \rangle\}^2 \rangle^{1/2}$), dashed lines represent those of the $y$- and $z$-component average ($\{\Delta v_y^2+\Delta v_z^2\}^{1/2}/2$).
A typical sound speed in molecular gas, $c_{\rm s}\simeq 0.3\,(T/20\,\mbox{K})^{1/2}$ km s$^{-1}$, is also plotted as a dashed line, where the temperature $T=20$ K is the CO-density-weighted average temperature at $t=10$ Myr.
In previous studies, it was shown that turbulence driven by the thermal instability (and the nonlinear thin shell instability) can induce ``supersonic turbulence" as a superposition of translative motions of clumps in the diffuse warm gas (Koyama \& Inutsuka 2002), but the cold clumps have subsonic internal velocity dispersions (Koyama \& Inutsuka 2006, Heitsch et al. 2006).
However, in the present simulation, as we will show in \S 3.5.1, the velocity dispersion can be supersonic in cold clumps depending on their size (see Figure \ref{f9}).
This qualitative difference can be attributed to the additional generation of turbulence triggered by the interaction between the accreting CNM and the accretion shock, i.e., the Richtmyer-Meshkov instability, which was absent in previous studies of cloud formation.

Figure~\ref{f7} clearly shows that the turbulence in the molecular cloud is anisotropic.
This is quite reasonable because the orientation of the converging flows in the simulation is fixed that leads to the selective input of the source of the turbulence (see also V\'azquez-Semadeni et al. 2007).
It is widely known that most observed molecular clouds are filamentary, indicating that the building blocks of molecular clouds are also accumulated anisotropically.
Thus, we can reasonably expect that turbulence in molecular clouds is anisotropic as well.
Recently, Hansen et al. (2011) reported that the dissipation rate of anisotropic turbulence can be smaller than that of isotropic turbulence, because the timescale of the cascade is determined by the crossing timescale in the direction of smaller velocity dispersion.
As we have discussed in the previous sections, the cold clumps are embedded in diffuse warm gas with typical sound speed $c_{\rm s}\gtrsim 5$ km s$^{-1}$, because the volume averaged temperature of the shock bounded region is $\langle T \rangle=2280$ K at $t=10.0$ Myr.
In isothermal gas, which is assumed in the conventional picture of molecular clouds, any supersonic motion of a cold clump creates a shock wave that decelerates the clump motion.
However, since the sound speed in the warm gas is larger than the velocity of the cold clumps ($\sim$ velocity dispersion of the turbulence), the molecular cloud formed in this simulation is a system with less shock waves than the conventional isothermal model in which the dissipation rate of turbulence is diminished.
Thus, owing to the effects of anisotropy and the diffuse warm gas, we can expect supersonic turbulence to be sustained longer, even after the accretion ceases.

\subsection{Magnetic Field}
\begin{figure}[t]
\epsscale{1.}
\plotone{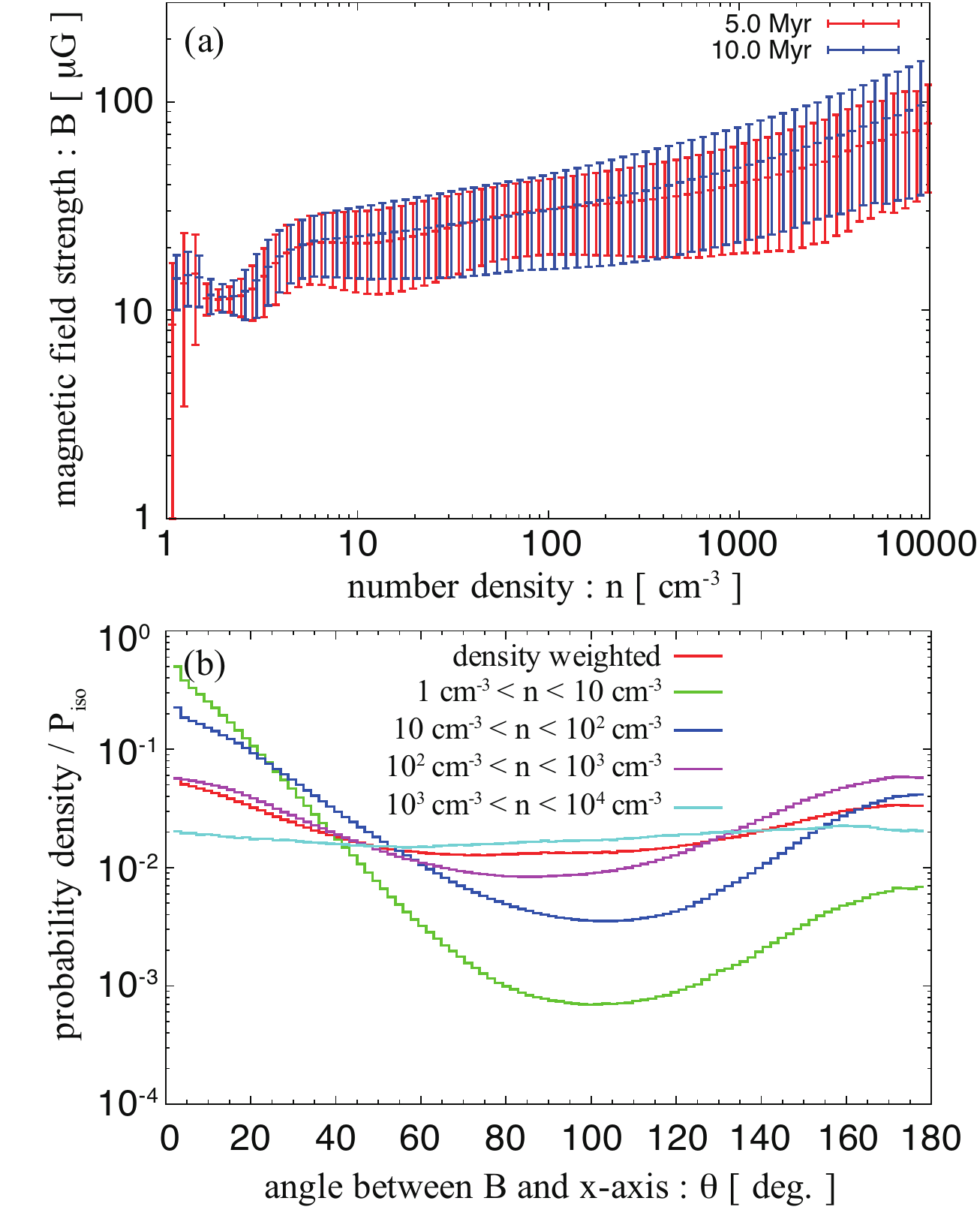}
\caption{
Panel (a): Average strength of magnetic field ($\langle |\vec{B}|\rangle$) as a function of density at times $t=5.0$ Myr ({\it red}) and $t=10.0$ Myr ({\it blue}).
Bars indicate the dispersion of the magnetic field strength at a given density.
Panel (b): Probability distribution function of the angle $\theta$ between the local direction of the magnetic field and the initial direction ($+x$ direction) in the shock bounded region at $t=10$ Myr.
In panel (b), the plotted functions are normalized by dividing the calculated distribution by the isotropic distribution $P_{\rm iso}(\theta)\propto\sin(\theta)$, i.e., {\it the flat distribution corresponds to an isotropic distribution}.
}
\label{f8}
\end{figure}

In the present simulation, we set up accretion flows along the mean magnetic field, because a molecular cloud can only be formed when the raw materials are accumulated along the mean magnetic field (Inoue \& Inutsuka 2009).
This leads to a constant mean magnetic field strength ($|\langle \vec{B} \rangle|=5.0\,\mu$G) throughout the evolution of the molecular cloud.
Thus, the mass-to-flux ratio ($\mu=M/\Phi=2\,\langle \rho \rangle v_{\rm cnv}\,t/|\langle \vec{B} \rangle|$) becomes larger than the critical value ($\mu_{\rm cr}=0.13\,G^{-1/2}$; Mouschovias \& Spitzer 1976) well before the formation of molecular cloud at $t_{\rm cr}\simeq 2\,{\rm Myr}\,(\langle n \rangle/5\,{\rm cm}^{-3})^{-1}\,(v_{\rm cnv}/20\,{\rm km s}^{-1})^{-1}\,(|\langle \vec{B} \rangle|/5\,\mu{\rm G})$.
Note that the mass-to-flux ratio of the initial two-phase medium in the domain is $\mu/\mu_{\rm cr}=0.27$.
Even though the mean field is constant, magnetic field lines are locally stretched and compressed due to turbulence which results in a local amplification of the magnetic field.
In panel (a) of Figure~\ref{f8}, we plot the average magnetic field strength ($\langle |\vec{B}|\rangle$) as a function of the local density at $t=5.0$ Myr ({\it red}) and $t=10.0$ Myr ({\it blue}), where bars indicate the dispersion of the magnetic field strength at that density.
Clearly, the average magnetic field strength $\langle |\vec{B}|\rangle$ in the dense gas is much larger than the mean magnetic field strength $|\langle \vec{B}\rangle|=5.0\,\mu$G, indicating that the magnetic energy of the molecular cloud shown in Figure~\ref{f6} is dominated by the energy contained in magnetosonic waves rather than the energy of the global mean field.

Because the turbulence in the formed molecular cloud is not only supersonic but also super-Alfv\'enic, it can easily bend and stretch the magnetic field lines.
If the turbulence was isotropic, the passive nature of the magnetic field would result in an isotropic distribution of local magnetic field orientations.
However, as shown in the previous section, the turbulence is anisotropic and enhanced in the direction of the accretion flows.
This leads to an anisotropic distribution of local magnetic field orientations.
In panel (b) of Figure~\ref{f8}, we show the probability distribution function of the angle $\theta$ between the local direction of the magnetic field and the initial direction ($+x$ direction) in the shock bounded region at $t=10$ Myr.
The red line represents the density-weighted distribution function, and lines in green, blue, magenta, and light blue represent the distribution functions in gas with $1\,{\rm cm}^{-3}<n\le10\,{\rm cm}^{-3}$, $10\,{\rm cm}^{-3}<n\le10^2\,{\rm cm}^{-3}$, $10^2\,{\rm cm}^{-3}<n\le10^3\,{\rm cm}^{-3}$, and $10^3\,{\rm cm}^{-3}<n\le10^4\,{\rm cm}^{-3}$, respectively.
In panel (b), the plotted functions are normalized by dividing the calculated distribution by the isotropic distribution $P_{\rm iso}(\theta)\propto \sin(\theta)$, i.e., the flat distribution corresponds to the isotropic distribution.

The direction of the local magnetic field is biased towards the direction of the converging flows, except in high-density regions with $n>10^3$ cm$^{-3}$.
The degree of anisotropy decreases with increasing density.
This is because as the gas evolves towards increasing densities due to the condensation of thermally unstable gas and/or the collision of clumps, the local flow direction, which affects the orientation of the local magnetic field, is deflected from the initial direction of the converging flows.
Note that the nearly isotropic distribution in dense gas with $n>10^3$ cm$^{-3}$ does not indicate the random orientation of magnetic field in each dense clumps.
It is noteworthy that, even in the low density regions with $1\,{\rm cm}^{-3}<n\le10\,{\rm cm}^{-3}$, the orientation of the magnetic field is flipped from the original direction ($\theta > 90$ deg.) with certain probability.
The flipped orientation of the magnetic field even in the low-density gas reflects the fact that the postshock diffuse gas (shocked WNM) is not in a state of ``laminar postshock flow", but in a state of ``anisotropic turbulence".

\subsection{Clump Statistics and Evolution}
In the formed cloud, molecular clumps evolve towards larger masses due to condensational accretion of the thermally unstable diffuse gas and collisional coalescence of clumps.
In this section, we present the statistics and evolution of physical quantities of the clumps.
In the following, we define a clump as a connected region with density is larger than a threshold value $n_{\rm th}$. 
In order to minimize the effects of numerical noise, we use the data for clumps that are resolved by more than 100 numerical cells.

\subsubsection{Plasma $\beta$, $l$-$\delta v$ relation, and mass-to-flux ratio}
Figure~\ref{f9} shows scatter plots of the plasma $\beta$ ($\equiv\langle 8\,\pi\,p/B^2 \rangle$) as a function of clump mass, and plots of the velocity dispersion as a function of clump scale.
Points in red, green, blue, and magenta represent the clumps identified using threshold densities of $n_{\rm th}=100$ cm$^{-3}$, 500 cm$^{-3}$, 2,000 cm$^{-3}$, and 10,000 cm$^{-3}$, respectively.
Panels (a) and (b) show the plasma $\beta$ at $t=5.0$ Myr and $t=10$ Myr and panels (c) and (d) are the size-velocity dispersion relation at $t=5.0$ Myr and $t=10$ Myr, respectively.
The size of the clumps is evaluated as $l=\sqrt{I_{1}/M}$, where $I_{1}$ is the largest eigenvalue of the inertial matrix and $M$ is the mass of the clump.
We also used the geometric mean of the three eigenvalues and the cubic root of the clump volume, but they did not change the results significantly.
In panels (c) and (d), the typical sound speed in molecular gas, $c_{\rm s}\simeq 0.3\,(T/20\,\mbox{K})^{1/2}$ km s$^{-1}$, is plotted as a thin dotted line, where the average temperatures in the clumps with $n\ge10^2,\,10^3,$ and $10^4$ cm$^{-3}$ are $42,\,19,$ and $13$ K, respectively.
The panels (a) and (b) indicate that the clumps evolve with time towards larger mass and that the number of high-density clumps increases with time, while the typical plasma $\beta$ stays at $\langle \beta \rangle \simeq 0.4 $ almost independent of clump density and time.
The size-velocity dispersion relation (panels [e] and [f]) also shows the universal law of $\delta v \simeq 1.5$ km s$^{-1}$ $(l/1\mbox{ pc})^{0.5}$, irrespective of both the mean density and time.
This relation is compatible with the well known size-line width relation of molecular clouds (Larson 1981, Heyer \& Brunt 2004), and consistent with the density independent relation observed in molecular cloud L1551 by Yoshida et al. (2010).

\subsubsection{Virial Parameters}
\begin{figure}[t]
\epsscale{0.9}
\plotone{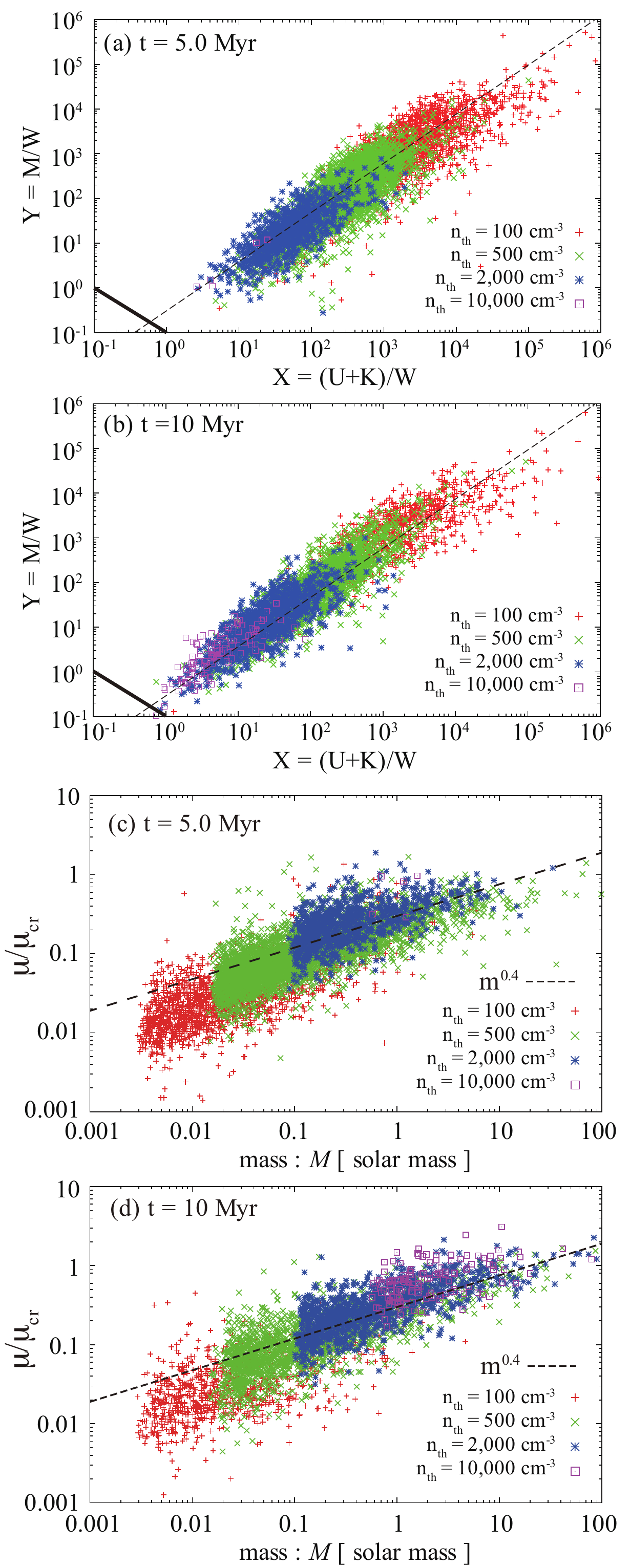}
\caption{
Panels (a) and (b): Scatter plots of clumps in a simple virial diagram in which horizontal and vertical axes represent $X\equiv (U+K)/W$ and $Y\equiv M/W$, respectively, where $U=\int 3\,p\,dV$, $K=\int \rho\,\delta v^2\,dV$, $M=\int B^2/8\pi\,dV$, and $W=\int \rho\,\vec{r}\cdot\vec{\nabla} \Phi\,dV$.
Panels (a) and (b) show the results at $t=5.0$ Myr and $t=10$ Myr, respectively.
A clump that enters the region of $X+Y<1$ (below the solid line) should evolve into star(s) by gravitational contraction.
Panels (c) and (d): Scatter plots of the mass-to-flux ratio as a function of clump mass at $t=5.0$ Myr and $t=10$ Myr, respectively.
}
\label{f10}
\end{figure}

As clumps evolve, their mass, density, and size increase towards becoming a gravitationally unstable object.
Panels (a) and (b) in Figure~\ref{f10} show scatter plots of the clumps in a simple virial diagram in which the horizontal and vertical axes represent $X\equiv (U+K)/W$ and $Y\equiv M/W$, respectively, where $U=\int 3\,p\,dV=3\,(\gamma-1)\,E_{\rm th}$, $K=\int \rho\,\delta v^2\,dV=2\,E_{\rm kin}$, $M=\int B^2/8\pi\,dV=E_{\rm mag}$, and $W=\int \rho\,\vec{r}\cdot\vec{\nabla} \psi\,dV=E_{\rm grv}$, where the post-process gravitational potential $\psi$ is used in the evaluation of the gravitational energy term $W$ (see \S 3.2).
Panels (a) and (b) show the results at $t=5.0$ Myr and $t=10.0$ Myr, respectively.
A clump that enters the region of $X+Y<1$ can evolve into star(s) due to gravitational contraction.
Because clumps grow for $\beta \simeq 0.4$ or $M\simeq U$ irrespective of their density, the clumps lie approximately along the line $X=Y$ for $X\gg1$.
For $X\sim1$, the kinetic energy term $K$ due to turbulence becomes comparable to or larger than the thermal energy term $U$.
Thus, the clumps become gravitationally unstable cores with $X\gtrsim Y$, indicating that magnetically super-critical cores (whose thermal and turbulent kinetic energies are larger than the magnetic energy) are formed as a consequence of the clump evolution.

Scatter plots of the mass-to-flux ratio ($\{E_{\rm grv}/E_{\rm mag}\}^{1/2}\simeq \mu/\mu_{\rm cr}$) as a function of clump mass at $t=5.0$ Myr and $t=10.0$ Myr are also shown in panels (c) and (d) of Figure~\ref{f10}, respectively.
Clumps identified with the same $n_{\rm th}$ follow $\mu/\mu_{\rm cr}\propto M^{0.4}$, which is consistent with the result of Banerjee et al. (2009), while a higher $n_{\rm th}$ leads to a larger mass-to-flux ratio for the same mass.
When collisional coalescence of clumps and accretion of thermally unstable gas onto the clumps occur along the magnetic field, the mass-to-flux ratio of clumps is enhanced.
Typical high density clumps with $n_{\rm th}=10^4$ cm$^{-3}$ can be super-critical when they grow to $M\gtrsim 1$ M$_{\rm solar}$.
It is widely known that the effects of ambipolar diffusion can enhance the mass-to-flux ratio even by condensational accretion of thermally unstable gas onto clumps perpendicular to the magnetic field (Inoue et al. 2007, Stone \& Zweibel 2010) and when clumps are shocked by collisional coalescence (Li \& Nakamura 2004, Kudoh \& Basu 2011).
Thus, if we include the effects of ambipolar diffusion, the clumps can be super-critical even for $M<1$ M$_{\rm solar}$.
Note that clumps far from the unstable state $X,\,Y\gg1$ (and $\mu/\mu_{\rm cr}\ll 1$) are small-scale, low-density, low-mass objects, and hence they should not be observed as dense magnetically subcritical cores with column density large enough to enable Zeeman effect observations (e.g., Crutcher et al. 2009).

\subsubsection{Rotation of clumps}
\begin{figure}[t]
\epsscale{0.9}
\plotone{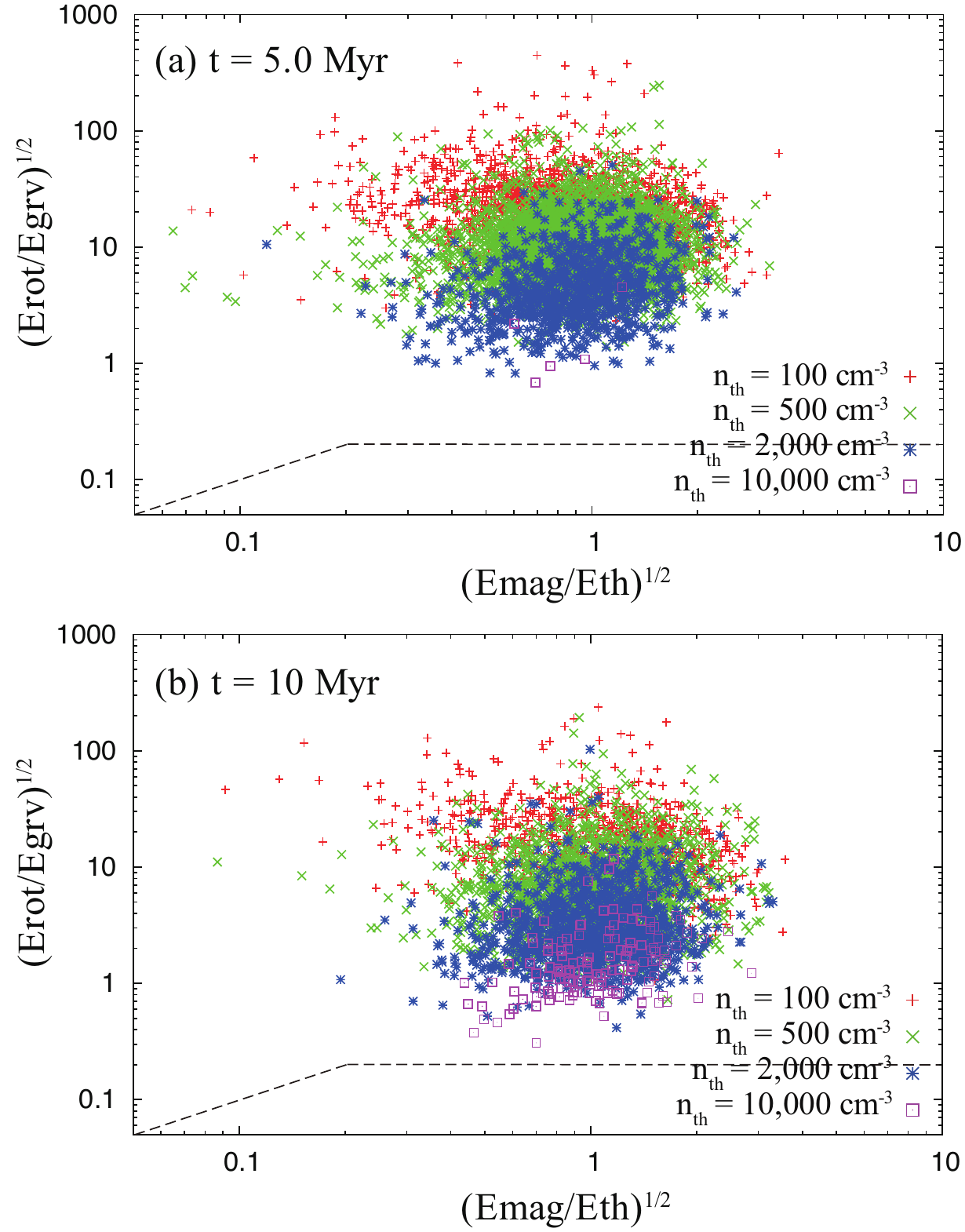}
\caption{
Scatter plot in the $(E_{\rm mag}/E_{\rm th})^{1/2}$-$(E_{\rm rot}/E_{\rm grv})^{1/2}$ plane.
Panels (a) and (b) show the results at $t=5.0$ Myr and $t=10.0$ Myr, respectively.
The dashed line represents the critical line of Machida et al. (2005) below which gravitational collapse leads to fragmentation into multiples.
}
\label{f11}
\end{figure}

Recent numerical simulations of the gravitational collapse of molecular cloud cores have revealed that cores can become fragmented after the firstcore is formed, if the initial condition of the core satisfies
\begin{eqnarray}\nonumber
 (E_{\rm rot}/E_{\rm grv})^{1/2}\gtrsim\left\{ \begin{array}{ll}
(E_{\rm mag}/E_{\rm th})^{1/2} & (\sqrt{E_{\rm mag}/E_{\rm th}}\lesssim0.2) \\
0.2 & (\sqrt{E_{\rm mag}/E_{\rm th}}\gtrsim0.2) \\
\end{array} \right.
\end{eqnarray}
where $E_{\rm rot},\,E_{\rm mag},\,E_{\rm th}$, and $E_{\rm grv}$ are the rotational, magnetic, thermal, and gravitational energies of the clump (Machida et al. 2004, 2008).
Figure~\ref{f11} shows a scatter plot in the $(E_{\rm mag}/E_{\rm th})^{1/2}$-$(E_{\rm rot}/E_{\rm grv})^{1/2}$ plane, where the solid line is the critical line for the above fragmentation condition.
Panels (a) and (b) show the results at $t=5.0$ Myr and $t=10.0$ Myr respectively.
Since even high-density threshold clumps with $n_{\rm th}=10^4$ cm$^{-3}$ that are close to the state of gravitationally bound cores are located in the fragmentation region (above the dashed line), most clumps seem to evolve into cores that satisfy the fragmentation condition.
The ratio of the rotational energy to the gravitational energy of observed molecular cloud cores is typically $\sim 0.2$ or $(E_{\rm rot}/E_{\rm grv})^{1/2}\sim 0.14$ (e.g., Goodman et al. 1993) that suggests non-fragmentation for cores of $E_{\rm mag}/E_{\rm th}\sim1$.
However, as shown in Dib et al. (2010), observations tend to underestimate the angular momentum by typically an order of magnitude which leads to comparable rotational energies between the observations and our simulation.
Note that, since neither the parameter $E_{\rm mag}/E_{\rm th}$ nor $E_{\rm rot}/E_{\rm grv}$ is a conserved variable, these parameters at the epoch of the beginning of the gravitational collapse are necessary to clarify whether the clumps truly evolve into cores that eventually fragment to form binaries, which requires a longer time simulation including the effect of self-gravity.

\subsubsection{Mass spectrum}
\begin{figure}[t]
\epsscale{1.}
\plotone{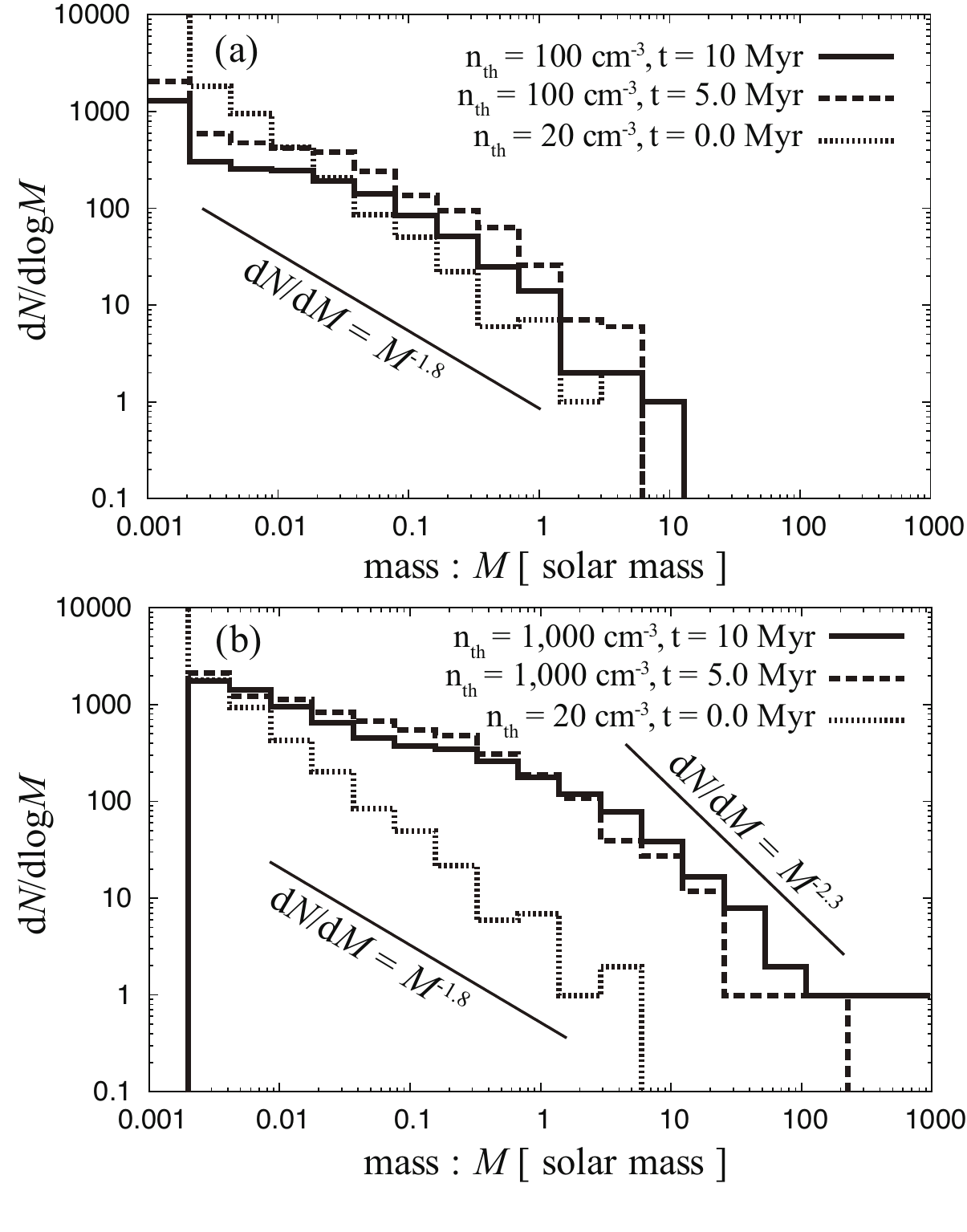}
\caption{
Mass spectrum of clumps at $t=5.0$ Myr ({\it solid}), and 10 Myr ({\it dashed}).
Panels (a) and (b) represent the spectra for $n_{\rm th}=100$ cm$^{-3}$ and 1,000 cm$^{-3}$, respectively.
The spectrum of initial H\textsc{i} clouds is also plotted in both panels as a dotted line.
}
\label{f12}
\end{figure}

Finally, we show the mass spectrum of the clumps.
We plot the mass spectrum of the clumps at $t=5.0$ Myr, and 10.0 Myr in Figure~\ref{f12}.
Panel (a) and (b) represent the spectra for $n_{\rm th}=100$ cm$^{-3}$ and 1000 cm$^{-3}$, respectively.
The mass spectrum of initial H\textsc{i} clouds are plotted as dotted lines in both panels, where the threshold density is set to be $n_{\rm th}=20$ cm$^{-3}$ (a few times smaller than the typical density of the initial H\textsc{i} clouds).
The mass spectrum of the initial H\textsc{i} clouds (dotted line) can be represented by a power law of the form $dN/dM\propto M^{-p}$ with $p\sim 1.8$.
This value of the spectral index $p$ can be understood from the theoretical mass spectrum of clumps formed by the thermal instability (Hennebelle \& Audit 2007) in which the spectral index is calculated as $p=2-(n-3)/3=1.78$, where $n=11/3$ is the spectral index of the seed fluctuation grown by the thermal instability that is used in the initial condition of H\textsc{i} cloud formation (see, \S2.2).
The spectra at later times show qualitatively different aspects depending on the threshold density.
The spectrum with a lower threshold density $n_{\rm th}=100$ cm$^{-3}$ (panel [a]) is similar to the spectrum of the initial H\textsc{i} clouds, which is consistent with observations of molecular clumps (Kramer et al. 1996, Schneider 2002).
At this lower threshold density, the spectra represent clumps generated by the compression of raw H\textsc{i} clouds by the two accretion shocks.
Thus, it is reasonable for the spectra in panel (a) to show similar shape to the initial H\textsc{i} cloud spectrum.
On the other hand, the higher density threshold clumps in panel (b) show very different spectra from the initial one at both $t=5.0$ and $10.0$ Myr.
Because the clumps identified using the high-density threshold would be more evolved, the deviation from the initial spectrum is not surprising.
The shallower slope of the spectra for $M\lesssim 1$ M$_{\rm sol}$ seems to suggest the growth of mass due to the coalescence of clumps by clump-clump collisions.
It is noteworthy that, for $M\gtrsim 1$ M$_{\rm sol}$, the spectra can be fitted by a power law with $p=2.3$, which resembles the core mass function in molecular clouds (e.g., Ikeda et al. 2007, Ikeda \& Kitamura 2009) and the high-mass tail of the initial mass function (Salpeter 1955), although the number of sample clumps is limited.

\section{Summary and Discussion}
Using three-dimensional magnetohydrodynamic simulations including the effects of radiative cooling/heating, chemical reactions, and thermal conductivity, we have investigated the formation of molecular clouds in the ISM.
We considered the formation of a molecular cloud due to the accretion of H\textsc{i} clouds formed by the thermal instability.
Such a mode of molecular cloud formation is compatible with the results of recent observations (Blitz et al. 2006, Fukui et al. 2009).
Since the mean density of the initial multiphase H\textsc{i} medium is an order of magnitude larger than the typical diffuse WNM density, the formation timescale is shorter than in the previous studies of molecular cloud formation that assumed the accumulation of only diffuse WNM.
The resulting timescale of molecular cloud formation of $\sim 10$ Myr is consistent with the evolutionary timescale of the molecular clouds in the LMC (Kawamura et al. 2009).

The results of our simulation can be summarized as follows:
\begin{itemize}
\item The initial multiphase H\textsc{i} medium is compressed and piled up behind shock waves induced by accretion flows.
Because the initial medium is highly inhomogeneous, the accretion shocks are highly deformed.
The deformed shock waves generate vortical motion behind the shock waves as a consequence of Crocco's theorem that results in the formation of a very turbulent molecular cloud.
\item The structure of the post shock region is roughly composed of dense cold gas ($T<100$ K) and diffuse warm gas ($T>1,000$ K), and these two components are spatially well mixed by the turbulence (\S 3.1).
Because the source of the turbulence is the accretion flows, the turbulence is highly anisotropic whose strength is biased toward the orientations of the accretion flows (\S 3.3).
\item The kinetic energy of the turbulence dominates the thermal, magnetic, and gravitational energies in the total 10 Myr evolution.
However, the kinetic energy measured by using the CO-fraction-weighted density is comparable to the other energies once the CO molecules are sufficiently formed owing to the UV shielding at $5$-$10$ Myr after the passage of shock wave, consistent with the fact that most of the molecular clouds are roughly virialized.
This suggests that the true kinetic energy of the turbulence in the cloud as a hole can be much larger than the kinetic energy of the turbulence estimated by using line-width of molecular emissions (\S 3.2).
\item Since the turbulence is highly super-Alfv\'enic, local magnetic field strength in dense gas is amplified due to the so-called turbulent dynamo in addition to the compression, while the mean magnetic field strength in the cloud as a hole stays at the initial value.
The local orientation of the magnetic field is also biased toward the direction of the accretion flows due to the effect of anisotropic turbulence, except the dense gas with $n>10^3$ cm$^{-3}$ (\S 3.4).
\item The clumps in the molecular cloud show statistically homogeneous evolution as follows:
(i) The typical plasma $\beta$ of the clumps is roughly constant $\langle \beta \rangle\simeq 0.4$;
(ii) The size-velocity dispersion relation show $\Delta v \simeq 1.5$ km s$^{-1}$ $(l/1\mbox{ pc})^{0.5}$ irrespective of density;
(iii) the clumps evolve towards magnetically supercritical cores (see, Figure~\ref{f10});
(iv) the clumps seem to evolve into gravitationally unstable cores that satisfy the fragmentation condition provided by Machida et al. (2005);
(v) the mass function of the low density clumps can be fitted by a power law $dN/dM\propto M^{-1.8}$ that reflects the initial spectrum of the H\textsc{i} clouds formed by the thermal instability, while that of the high density clumps suggests steeper slope of $dN/dM\propto M^{-2.3}$ for the high mass tail.
\end{itemize}

In a subsequent paper, using synthetic observations of polarized dust emission, we will show that the anisotropic orientation of magnetic field leads to an overestimation of the mean magnetic field strength when employing the Chandrasekhar-Fermi effect (Chandrasekhar \& Fermi 1953).
In this paper, we have omitted the dynamical effects of self-gravity.
For the duration of our simulation ($t=10$ Myr), the neglect of self-gravity can be justified, because the gravitational energy has not yet became the dominant component of the energy budget in the molecular cloud (see Figure~\ref{f6}) and almost all dense clumps in the cloud are still in the gravitationally unbound state (see Figure~\ref{f10}).
However, as clearly seen in Figure~\ref{f6}, the gravitational energy increases continuously, while other energies plateau.
Thus, in order to understand the later time evolution, the inclusion of self-gravity is crucial (see Hennebelle et al. 2008, Heitsch et al. 2008, Banerjee et al. 2009, Vazquez-Semadeni et al. 2011 for the cases of colliding WNM flows), which we will study in our next paper.

\acknowledgments
We would like to thank Kazuyuki Omukai, Takashi Hosokawa, Kohji Tomisaka, and Jennifer M. Stone for fruitful discussions.
Numerical computations were carried out on XT4 at the Center for Computational Astrophysics (CfCA) of National Astronomical Observatory of Japan.
This work is supported by Grant-in-aids from the Ministry of Education, Culture, Sports, Science, and Technology (MEXT) of Japan, No. 21740146, No.22$\cdot$3369 and No. 23740154 (T. I.), and No. 16077202 and No. 18540238 (S. I.).

\clearpage

\end{document}